\documentclass[conf]{new-aiaa}
\usepackage[utf8]{inputenc}
\usepackage{subcaption}
\usepackage{amsmath}
\usepackage{graphicx}
\usepackage[version=4]{mhchem}
\usepackage{siunitx}
\usepackage{bm}
\usepackage{longtable,tabularx}
\usepackage{booktabs}
\usepackage{algorithm} 
\usepackage{algpseudocode}
\usepackage{enumitem}
\usepackage{tikz}
\usepackage{ulem}

\newif\ifcomments
\commentsfalse
\commentstrue 
\ifcomments
\newcommand{\mmm}[1]{{\color{red}\textbf{MM: #1}}}      
\newcommand{\pjj}[1]{{\color{magenta}\textbf{PJ: #1}}}    
\newcommand{\lc}[1]{{\color{blue}\textbf{LC: #1}}}    
\else
\newcommand{\mmm}[1]{}
\newcommand{\eag}[1]{}
\newcommand{\pjj}[1]{}
\newcommand{\lc}[1]{}
\fi

\definecolor{myblue}{RGB}{0,154,249}
\definecolor{mygreen}{RGB}{0,128,0}
\definecolor{myorange}{RGB}{226,111,70}
\definecolor{mymagenta}{RGB}{195,113,211}

\newcommand{\mysquare}{
\begin{tikzpicture}[scale=1.5, transform shape]
    \fill[color = myblue] (0,0) rectangle (0.2,0.2);
\end{tikzpicture}
}
\newcommand{\mytriang}{
\begin{tikzpicture}[scale=1.5, transform shape]
    \fill[color = mygreen] (0,0) -- (0.1,0.2) -- (0.2,0) -- cycle;
\end{tikzpicture}
}
\newcommand{\myitriang}{
\begin{tikzpicture}[scale=1.3, transform shape]
    \fill[color = mymagenta] (0.1,0) -- (0.0,0.2) -- (0.2,0.2) -- cycle;
\end{tikzpicture}
}
\newcommand{\mycirc}{
\begin{tikzpicture}[scale=1.5, transform shape]
    \fill[color = myorange] (0.1,0.1) circle (0.1);
\end{tikzpicture}
}

\newcommand{\dd}{\mathrm{d}}
\newcommand{\pd}{\partial}

\title{An implicit, conservative electrostatic particle-in-cell algorithm for paraxial magnetic nozzles}
\author{Pedro Jim\'enez}
\affil{Equipo de Propulsi\'on Espacial y Plasmas (EP2), Universidad Carlos III de Madrid, Legan\'es, Spain}  
\author{Luis Chac\'on}
\affil{Los Alamos National Laboratory, Los Alamos, NM, USA} 
\author{Mario Merino}
\affil{Equipo de Propulsi\'on Espacial y Plasmas (EP2), Universidad Carlos III de Madrid, Legan\'es, Spain} 

\begin{document}

\maketitle

%
\begin{abstract}
An electrostatic, implicit particle-in-cell (PIC) model
for collisionless, fully magnetized, 
paraxial plasma expansions in a magnetic nozzle is introduced with exact charge, energy, and magnetic moment conservation properties. 
The approach is adaptive in configuration space by the use of mapped meshes, and exploits the strict conservation of the magnetic moment to reduce the dimensionality of velocity space.
A new particle integrator is implemented, which allows for particle substepping without the need to stop particle motion at every cell for charge conservation. Particle suborbits are determined from accuracy considerations, and are allowed to span multiple cells.
Novel particle injection and expansion-to-infinity boundary conditions are developed, including a control loop to prevent the formation of spurious sheaths at the edges of the domain.
The algorithm is verified in a periodic magnetic mirror configuration, a uniform plasma test case (to test particle injection), and a propulsive magnetic nozzle. 
The algorithm's computational complexity is shown to scale favorably with timestep, and  linearly with the number of particles and mesh cells (unlike earlier implicit PIC implementations, which scaled quadratically with the number of mesh cells in one dimension). Numerical experiments demonstrate that the proposed algorithm outperforms both explicit PIC and semi-Lagrangian Vlasov codes by more than an order of magnitude.  
\end{abstract}
%

\section{Introduction}

A guiding divergent magnetic field can be used to expand a plasma in a controlled and directed manner to generate thrust. This is the principle behind the magnetic nozzle (MN) \cite{ande69,ahed10f}, a device used to convert plasma thermal energy into macroscopic kinetic energy contactlessly, which constitutes an essential 
component of several plasma propulsion technologies such as the 
helicon plasma thruster \cite{char03b,ziem05,nava18a}, the electron cyclotron resonance thruster \cite{kuni98,vial18,svil21a}, the applied-field magnetoplasmadynamic thruster \cite{saso92, krul98}, and the variable specific impulse magnetoplasma rocket \cite{diaz00}.
In contrast to a classical (material) de Laval nozzle, where the pressure force on the device walls is the mechanism limiting the radial expansion of the gas and leading to the generation of mechanical thrust, in a MN it is the Lorentz force on the charged particles that confines their radial expansion and generates magnetic thrust. 

The expansion in the MN is quasi-collisionless, which limits the applicability of closed fluid models.  
For this reason, kinetic models have been used to understand, in particular, the electron response, which is intimately linked to the development of the ambipolar electrostatic potential responsible for the acceleration of the ions.
Martínez-Sánchez et al. \cite{mart15a} established a steady-state, semi-analytical  paraxial model of the expansion of an initially Maxwellian plasma in a converging-diverging magnetic field. They 
showed the presence of different electron subpopulations (free streaming, reflected, and doubly-trapped electrons) depending on the connectivity of their trajectories with the upstream plasma source and infinity downstream, 
as determined by barriers of an effective potential that results from the electrostatic field and the magnetic mirror force.
That study revealed the existence of electron cooling and electron anisotropization on the divergent side, different for each electron subpopulation. The overall electron species exhibits a behavior that corresponds to the weighted average of the three subpopulations.
While that model solves for the distribution function of ions and electrons whose trajectories connect with the boundary conditions of the problem, it is unable to resolve the distribution of doubly trapped electrons, which was hypothesized to coincide with the distribution upstream.
The model was later used to explore approximate closure relations \cite{ahed20a}, and was extended to two-dimensional MNs \cite{meri21a}.

Sánchez et al. \cite{sanc18b}
explored the problem of electron kinetics in a MN with a time-dependent Vlasov code, and the same team later added weak resistivity on the electron dynamics \cite{zhou21a}. These two mechanisms enabled the self-consistent study of doubly trapped electrons, as these regions are populated during the transient plume set-up and by collisional
diffusion.
However, in addition to the high computational cost of this kinetic code, numerical difficulties were identified in the definition of the upstream and downstream boundary conditions (BCs). The choice of BCs in those studies led to the formation of undesired Debye sheaths in simulations that do not correspond with the expansion of a MN plasma into free space. This forced the truncation of the analysis domain to about half of the simulation domain. In other studies, the open-boundary BCs used prevented simulation of steady-state plasma expansion, as the time evolution of the plume had to be stopped before the particles reached the end of the domain \cite{hu17,brie18}.

Advances in the proper treatment of BCs in kinetic simulations were presented by Li et al. \cite{li19a}. That study showed that, when using Dirichlet conditions for the potential at the entrance and homogeneous Neumann conditions at the exit, if all electrons reaching the downstream boundary are removed from the simulation, a numerical instability develops in the whole domain. Rather, a more consistent approach is to acknowledge that most of these electrons do not have the mechanical energy to reach infinity downstream and must be reflected based on the value of the total potential fall $\phi_\infty-\phi_0$, a magnitude that must be determined self-consistently during the simulation to enforce global current ambipolarity in the plasma jet.
However, the reference employed an explicit particle-in-cell (PIC) approach, which was limited to very small domains due to computational constraints.  Andrews et al. proposed yet another strategy based on a multipole expansion that yields non-stationary Robin-type boundary conditions on Poisson's equation \cite{andr22}.  
 
Recently, a class of implicit electrostatic PIC algorithms have been proposed \cite{chen11} that are exactly charge- and energy-conserving, and able to leverage adaptive meshing without loss of those conservation properties \cite{chacon13map}. 
These algorithms have been extended to deal efficiently with low-frequency electromagnetic regimes \cite{chen14em,chen15em} and strongly magnetized regimes \cite{chen22}. 
The ability of these schemes to use time steps much larger than plasma and cyclotron frequencies, along with the preservation of strict conservation properties, makes them particularly attractive for the simulation of MNs in multiple dimensions and arbitrary magnetization regimes.

In this study, we take the first step along this direction by exploring the use of implicit electrostatic PIC algorithms for paraxial magnetic nozzles in the strongly magnetized regime. We build on earlier studies and adapt a novel particle mover (particularly suitable for strongly magnetized regimes \cite{chen22}) to mapped adaptive meshes. 
Moreover, a dynamic outflow electric-field boundary condition in one spatial dimension
is rigorously derived 
that captures the plasma expansion towards infinity without introducing numerical artifacts (such as artificial sheaths) or forcing arbitrary conditions on the electric field. We demonstrate the new implicit 1D PIC solver on several examples including a magnetic mirror, a uniform plasma, and a paraxial magnetic nozzle. For the latter, we compare our results with published data \cite{sanc18b}, finding a good agreement.
We show that our model avoids domain-termination sheaths and outperforms \cite{sanc18b} in accuracy and computational time.

The rest of the paper is structured as follows: Section \ref{sec:mod} introduces the quasi 1D paraxial model, field equations and the particle discretization of phase space. Section \ref{sec:num} tackles the numerical discretization and solution of the model via an implicit particle in cell algorithm, including the generalization of the new segment-averaged mover \cite{chen20a} to mapped geometries. It also proves the global energy and local charge conservation properties of the method. Section \ref{sec:bound} introduces the newly developed injection algorithm and dynamic downstream boundary condition for MNs.
Section \ref{sec:verification} discusses the verification of the code with three numerical examples. The first (\ref{sec:mirror}) demonstrates conservation properties and the effect of the magnetic mirror force in a simplified periodic geometry; the second (\ref{sec:uniformplasma}) demonstrates the ability of the code to handle particle injection in an implicit PIC context with a uniform plasma verification example; and the third (\ref{sec:nozzle}) compares the full MN model against literature results. Finally, Section \ref{sec:concl} presents a summary of this work.

\section{Model}\label{sec:mod}

The kinetic model describes the collisionless, quasineutral expansion of a plasma from an upstream source into vacuum along the axis of an axisymmetric, convergent-divergent magnetic field $\bm{B}$, known as magnetic nozzle (MN). The maximum $B$ is located at $z=0$ and is referred to as magnetic throat.
The plasma is composed of strongly-magnetized electrons $e$ and singly-charged ions $i$ that satisfy 
\begin{align}
\frac{m_sv_s}{q_s B} &\ll L_r \ll L_z,
&
\frac{m_s}{q_s B} &\ll \tau,
\label{eq:paraxiallity}
\end{align}
for $s=i,e$,
where $L_z$, $L_r$ are the characteristic gradient lengths in the axial and radial directions and $\tau$ the characteristic time of change of the electric and magnetic fields and the plasma distribution functions. 

\subsection{Paraxial drift-kinetic equation} 

Under the assumptions above, the magnetic moment of each charged particle 
$\mu_s= m_s v_{\perp}^2 / (2B)$ 
is a conserved adiabatic invariant, and
the gyroaveraged 
distributions $f_s(\mathbf{x},v_\parallel,\tilde \mu,t)$  
($s=i,e$) (with $\Tilde{\mu}= \mu_s/m_s$ the mass-scaled magnetic moment) along the axis 
respond to the drift-kinetic Vlasov equation (DKE) in conservative form \cite{hazeltine2003plasma}:
\begin{equation}
\frac{\pd f_s}{\pd t} 
+ \frac{1}{B(\mathbf{x})} \left [ \nabla \cdot \left ( v_\parallel \mathbf{B} f_s \right )
+ \frac{\partial}{\partial v_\parallel} \left( \left ( \frac{q_s}{m_s} E_\parallel - \Tilde{\mu} \mathbf{b}\cdot\nabla B \right)  B f_s \right ) \right ] 
= 0,
\label{eq:cons-vlasov}
\end{equation}
where $E_\parallel=-\mathbf{b}\cdot \nabla \phi$ is the parallel electrostatic field, with $\phi(\mathbf{x},t)$ the electrostatic potential. Perpendicular drifts have been neglected, as they are strictly zero in the paraxial approximation (see below). The factor $B(\mathbf{x})$ in this equation is the Jacobian of the transformation from $(\mathbf{x},v_\parallel,v_\perp^2/2)$ to 
$(\mathbf{x},v_\parallel,\Tilde{\mu})$ coordinates:
\begin{equation}
    J_\mu=\frac{\partial(\mathbf{x},v_\parallel,v_\perp^2/2)}{\partial(\mathbf{x},v_\parallel,\Tilde{\mu})}=B(\mathbf{x}).
\end{equation}
Noting that the phase-space flow $\left ( v_\parallel \mathbf{B},B(\frac{q_s}{m_s} E_\parallel - \Tilde{\mu} \mathbf{b}\cdot\nabla B) \right )$ is incompressible (as required for the conservation of phase-space volume),
\begin{equation}
    \nabla \cdot \left ( v_\parallel \mathbf{B} \right )
+ \frac{\partial}{\partial v_\parallel} \left ( B \left ( \frac{q_s}{m_s} E_\parallel - \Tilde{\mu} \mathbf{b}\cdot\nabla B \right ) \right ) =0,\label{eq:ps_incompressibility}
\end{equation}
%
the DKE can be rewritten in characteristic form as:
\begin{equation}
\partial_{t}f_{s}+v_{\parallel}\mathbf{b}\cdot\nabla f_{s}+\left(\frac{q_{s}}{m_{s}}E_{\parallel}-\tilde{\mu}\mathbf{b}\cdot\nabla B\right)\frac{\partial f_{s}}{\partial v_{\parallel}}=0.\label{eq:rDKE-nc}
\end{equation}

In the paraxial approximation, we consider a 1D representation along
the axis $z$,
and have:
\[
v_{\parallel}\approx v_{z}\,\,;\,\,
E_{\parallel}\approx E_{z}\,\,;\,\,
\mathbf{B}\approx B(z)\mathbf{z}.
\]
To recover the solenoidal property of the magnetic
field (and phase-space incompressibility),
we require the paraxial Jacobian
factor $J^{B}=1/B(z)$ in the definition of the divergence, such that:
\[
\nabla\cdot\mathbf{B}=\frac{1}{J^{B}}\partial_{z}(J^{B}B)=0.
\]
This Jacobian is proportional to the flux-tube area by flux conservation, i.e., $A_\mathrm{ft}(z)\propto 1 /B(z)=J^B$. The characteristic form of the paraxial DKE equation reads (from
Eq. \ref{eq:rDKE-nc}):
\begin{equation}
\partial_{t}f_{s}+v_{z}\partial_{z}f_{s}+\left(\frac{q_{s}}{m_{s}}E_{z}-\tilde{\mu}\partial_{z}B\right)\frac{\partial f_{s}}{\partial v_{z}}=0,\label{eq:pDKE-nc}
\end{equation}
with the corresponding conservative form (noting that $J_\mu J^B=1$):
\begin{equation}
\partial_{t}f_{s}+\frac{\partial}{\partial z}\left(v_{z}{f_{s}}\right)+\frac{\partial}{\partial v_{z}}\left(\left(\frac{q_{s}}{m_{s}}E_{z}-\tilde{\mu}\partial_{z}B\right){f_{s}}\right)=0.\label{eq:pDKE-cs}
\end{equation}

The corresponding particle representation of $f_s$ is given by:
\begin{align}\label{eq:pcle_ansatz}
    f_s(z,v_z,\tilde \mu,t) = \sum_{p \in s} w_p \delta(z-z_p(t)) \delta(v_z-v_{z,p}(t)) \delta(\tilde \mu - \tilde{\mu}_p(t)),
\end{align}
where $w_p$ is the weight of the $p$-th particle of species $s$.
Introducing the particle ansatz \eqref{eq:pcle_ansatz} into the Vlasov equation \eqref{eq:pDKE-nc} yields the axial particle motion governing equations:

\begin{align}
\frac{\dd w_p}{\dd t}&= 0,
\label{eq:pclewgt}
\\\frac{\dd \tilde{\mu}_p}{\dd t}&= 0,
\label{eq:pclemu}
\\
\frac{\dd z_p}{\dd t}&= v_{z,p},
\label{eq:pclepos}
\\
\frac{\dd v_{z,p}}{\dd t}&=\frac{q_p}{m_p} E_z(z_p) -  \tilde{\mu}_p\left.\frac{\dd B}{\dd z} \right |_{z_p}.
\label{eq:acc}
\end{align}

Zeroth and first moments of $f_s$ give charge and current densities affected by the paraxial Jacobian (as required  \cite{chac13} since particles carry charges and currents, not densities), and are given by:
\begin{align}
    J^B \rho(z,t) &= J^B \sum_s q_s \int dv_z d\tilde{\mu} J_\mu ~f_s(z,v_z,\mu,t) = \sum_{p} q_p w_p \delta(z-z_p(t)),  \\
    J^B j_z(z,t) &= J^B \sum_s q_s \int dv_z d\tilde{\mu} J_\mu ~v_z f_s(z,v_z,\mu,t) = \sum_{p} q_p w_p v_{z,p} (t)\delta(z-z_p(t)) .
\end{align}

\subsection{Electrostatic field equation}

The particle system is closed with an evolution equation for the electrostatic field. Instead of using Poisson's equation directly, as is customary in classical PIC algorithms, and with the goal of achieving strict numerical energy conservation in mind, we begin by considering Ampere's law \cite{chen11},
\begin{align}\label{eq: divAmp}
    \epsilon_0\frac{\partial\mathbf{E}}{\partial t}+\mathbf{j}=\nabla \times \mathbf{B}.
\end{align}
Taking the divergence of this equation gives:
\begin{align}
    \epsilon_0\frac{\partial}{\partial t} \nabla\cdot\mathbf{E}+ \nabla\cdot\mathbf{j} = 0.
\end{align}
Introducing the scalar electric potential $\mathbf{E}=-\nabla\phi$, one arrives at the final field equation used in this study:
\begin{align}\label{eq:ampere}
    \epsilon_0\frac{\partial}{\partial t} \nabla^2\phi = \nabla\cdot\mathbf{j}.
\end{align}
Note that this equation recovers the charge conservation equation if one replaces the Poisson equation or, equivalently, the same equation can also be derived by temporally differentiating Poisson's equation and using conservation of charge. 
Equations \eqref{eq:pDKE-nc} and \eqref{eq:ampere} for $f_s$ and $\phi$ must be complemented with the necessary initial  and boundary conditions.
In particular, at the upstream plasma source, the one-sided part of $f_s$ ($s=i,e$) with $v_z>0$ is assumed to be known. In our implementation, we do not use Eq. \ref{eq:ampere}, but instead we solve:
\begin{align}\label{eq:ampere-inv}
    \epsilon_0\frac{\partial \phi}{\partial t}  = \nabla^{-2}(\nabla\cdot\mathbf{j}),
\end{align}
which removes the numerical stiffness associated with the Laplacian operator. The discrete Laplacian operator in one dimension is a simple tridiagonal matrix, of straightforward inversion.

It is worth observing that the axial momentum equation \eqref{eq:acc} can be expressed in terms of the effective potential $U_{\mathrm{eff},s}$, defined in \cite{mart15a} as 
\begin{align}\label{eq:effpot}
U_{\mathrm{eff},s}(z, t; \tilde{\mu})
= \frac{q_s}{m_s}\phi(z,t) + \tilde{\mu} B(z),
\end{align}
for $s=i,e$, so $\dd v_{z,p}/\dd t = - \dd U_{\mathrm{eff},s}/\dd z |_{z_p}$. 
At every instant of time, while $\phi$ is monotonically decreasing along the MN in the cases of interest,
$U_{\mathrm{eff},s}$ is generally nonmonotonic, at least for some $\tilde \mu$. 
This creates potential barriers that cause the distributions of ions and electrons to subdivide into free-streaming, reflected, and trapped subpopulations, depending on 
the connectivity of their trajectories with the upstream plasma source 
and infinity downstream \cite{mart15a, ahed20a, meri21a}. 
These subpopulations are cleanly delimited in steady state, where $\phi$ only depends on $z$.
Naturally, in a time-dependent context, $U_{\mathrm{eff},s}$ varies in time, and the boundaries between these sub-populations become dynamic. Hence, initially free particles may become trapped during the transient setup of the plasma beam \cite{sanc18b}, and vice versa, as $\phi$ evolves and their mechanical energy changes. 

\subsection{Transformation of the spatial coordinate}

At this point, and for the purpose of spatial adaptivity, it is useful to consider analytically deformed meshes using a mesh map $z(\xi)$. In 1D, the Jacobian of the transformation is simply $J_{\xi}=\dd z / \dd \xi$. To ensure strict numerical energy conservation, we employ a hybrid push \cite{chac13,swif96,wang99} that evolves the particle position in logical space $\xi$, while the particle velocity is evolved in the Cartesian coordinate $v_z$. Following equation \eqref{eq:pclemu} $\tilde{\mu}_p$ is exactly conserved and there is no need to evolve it. Additionally, 
this approach recovers the  advantage of fast particle mesh localization and mesh-particle interpolations in a structured-mesh computational space, while avoiding cumbersome inertial forces in the particle equation of motion due to the presence of the map \cite{fichtl2012arbitrary}. The evolution of the particle position in logical space $\xi_p$ is found by dividing equation \ref{eq:pclepos} by $J_{\xi}$ to find: 
\begin{align}
\label{eq:xicont}
\frac{\dd \xi_p}{\dd t} = \frac{v_{z,p}}{J_{\xi}}. 
\end{align}
Regarding the momentum equation \eqref{eq:acc}, we similarly write the Cartesian electric field as $E_z = -\partial \phi / \partial z =  E_{\xi} / J_{\xi}$, with $E_{\xi}  = - \partial \phi / \partial \xi $ (which is directly computable on the logical mesh), to find: 
\begin{align}\label{eq:vcont}
    \frac{\dd v_{pz}}{\dd t} = -\frac{q_p}{m_p} \left( \frac{E_{\xi} }{J_{\xi}} \right )_{\xi_p} -  \frac{\mu_p}{m_p} \left.\frac{\dd B}{\dd z} \right |_{\xi_p}.
\end{align}

The logical representation of the field equation Eq. \eqref{eq:ampere} in 1D reads:
\begin{align}\label{eq:amperecurv}
    \epsilon_0 \partial_t \partial_\xi\left(J g^{\xi \xi} \partial_\xi \phi\right)=\partial_\xi\left(J j^\xi\right),
\end{align}
where $\partial_\xi = \partial / \partial \xi$. 
The Jacobian $J=J_\xi J^B$ includes both the mesh map contribution $J_\xi$ and the paraxial Jacobian $J^B=1/B(z)$.
In 1D, the contravariant metric tensor has just one component given by $g^{\xi \xi}=(J_{\xi})^{-2}$, and the contravariant current density vector reduces to just one component, $j^{\xi}=\mathbf{j}\cdot \nabla \xi = j_z / J_{\xi}$. Equation \eqref{eq:amperecurv}, in 1D logical coordinates $\xi$ and in the paraxial approximation, finally reads:
\begin{align}
    \epsilon_0
    \partial_t\partial_\xi\left(\frac{J^B}{J_{\xi}}\partial_\xi\phi\right) = \partial_\xi(J j^{\xi}),\label{eq:1d_ampere_curv}
\end{align}
where the contravariant current component is found from the Cartesian particle velocity component $v_{z,p}$ as in \cite{chac13}:
\begin{align}
    J j^{\xi} = \sum_p q_p \frac{v_{z,p}}{J_{\xi}} \delta (\xi - \xi_p(t)) = \sum_p q_p \frac{\dd \xi_p}{\dd t} \delta (\xi - \xi_p(t)).
\end{align}

\section{Numerical Implementation}\label{sec:num}
\subsection{Particle enslavement and subcycling}

After a suitable discretization, one may use Eqs. \eqref{eq:xicont} and \eqref{eq:vcont} with the new particle coordinates $\mathbf{X_2} =\left\{\xi^{n+1},v_z^{n+1}\right\}_p$ and Eq. \eqref{eq:amperecurv} for the new potential at the cell centers 
$\mathbf{X_1} = \left\{\phi^{n+1}_i\right\}$ to obtain a nonlinear residual vector $F(\mathbf{X_1},\mathbf{X_2})=0$, which can be solved iteratively. However, such a formulation renders the number of unknowns too large for practical deployment in current massively parallel computers. 

A practical implementation can be realized by noticing that the new particle coordinates are themselves functions of the potential, from which it follows that a new residual can be written $\mathbf{F}_1\left(\mathbf{X}_1, \mathbf{f}_2\left(\mathbf{X}_1\right)\right)=\mathbf{G}\left(\mathbf{X}_1\right)=0$ \cite{chen11}. Finding $\mathbf{f}_2\left(\mathbf{X}_1\right)$ requires an orbit integral, that is, for each particle, and given the electric potential on the mesh nodes $\mathbf{X_1}=\phi^{n+1}_i$, the equations of motion need to be integrated to accumulate the moments to form the residual vector $\mathbf{G}(\phi_i^{n+1})$. This procedure, known as \textit{particle enslavement} \cite{chen11}, allows a global nonlinear solver, in our case a Jacobian Free Newton Krylov (JFNK) method, to handle a much reduced system of equations with no compromise on the accuracy of the solution.

Particle enslavement in an implicit PIC context provides great versatility for the integration of particle orbits. In particular, particle sub-stepping or subcycling will be employed \cite{chen11}. Instead of using the overall time step $\Delta t$ to integrate the particle orbits numerically, different $\Delta \tau_p^\nu$ per substep $\nu$ (suitably determined, as discussed below) are used such that $\sum\Delta\tau_p^\nu = \Delta t$.
This approach can leverage the difference of scales between the slow field dynamics, characterized by $\Delta t$, and the fast particle dynamics, represented by $\Delta\tau_\nu^p$, to allow for accurate orbit integrals while avoiding expensive calls to memory because the particle coordinates are stored in local registers during the computation of subsequent substeps.

In earlier implicit PIC implementations \cite{chen11}, the simultaneous enforcement of local charge and energy conservation is achieved by adjusting intermediate $\Delta\tau_p^\nu$ to make particles stop at cell faces. Consequently, the number of cell crossings, and thus of substeps, grows proportionally to $\Delta t$ and scales with the mesh resolution (proportionally to the number of mesh points in 1D \cite{chen14}). In practice, this hinders efficiency gains from the use of large $\Delta t$. Recently, a new particle mover has been proposed \cite{chen22} that allows particles to cross several cells in a single substep while still conserving global energy and local charge exactly. The present work generalizes the new mover to nonuniform mapped meshes.


\subsection{Particle orbit integration}

We describe next the discretization of the particle equations of motion, the timestep estimator, and the orbit integration algorithm in mapped geometries. A key innovation of our approach is the ability of particles to traverse multiple cells in a single substep without spoiling conservation properties. 

\subsubsection{Crank-Nicolson discretization}

A fully implicit, time centered, second-order and non-dissipative Crank-Nicolson discretization \cite{chen11} is used,
\begin{align}
    &\frac{\xi_p^{\nu+1}-\xi_p^{\nu}}{\Delta\tau_p^\nu}= \left ( \frac{v_z}{J_{\xi}} \right)_p^{\nu+1 / 2}, \label{eq:xiCN}\\ 
    &\frac{v_{z,p}^{\nu+1}-v_{z,p}^{\nu}}{\Delta\tau_p^\nu}
    =
    a^{v+1/2}
    =\frac{q_{p}}{m_{p}} \left( \frac{E_{\xi} }{J_{\xi}} \right)^{\nu+1/2}_p - \Tilde{\mu}_p\frac{B_p^{\nu+1} - B_p^{\nu}}{v_{z,p}^{n+1/2}\Delta\tau_p^\nu},\label{eq:vCN}
\end{align}
where $v_{z,p}^{\nu+1 / 2}=(v_{z,p}^{\nu+1}+v_{z,p}^{\nu})/2$,
and $a^{\nu+1 / 2}$ is the particle acceleration computed at ($\Delta\tau/2$, $(\xi_{p}^{\nu+1}+\xi_{p}^{\nu})/2$)
from the electric and magnetic mirror forces. The definition of the Jacobian average over the substep, $J_\xi^{\nu+1 / 2}$, will account for multiple cell-crossings per particle substep, and is reported later in this section (Sec. \ref{sec:orbit-int}). 
The discretization proposed for the mirror force term, $\tilde{\mu}_p(\dd B/\dd z)$ in the right hand side of Eq. \eqref{eq:vCN}, has already used Eq. \ref{eq:xiCN} and will be shown to conserve energy exactly.

The coupling between $\xi_p^{\nu+1}$ and $v_{z,p}^{\nu+1/2}$ via the mid-orbit electric field $E_{\xi,p}^{\nu+1/2}$ implies that an additional non-linear solve per particle is needed inside the mover. The system of equations is well posed for large time-steps, and simple Picard iterations achieve convergence to a very tight tolerance in a few steps \cite{kosh22}.

\begin{figure}[t]
\centering
    \includegraphics[width=1\textwidth]{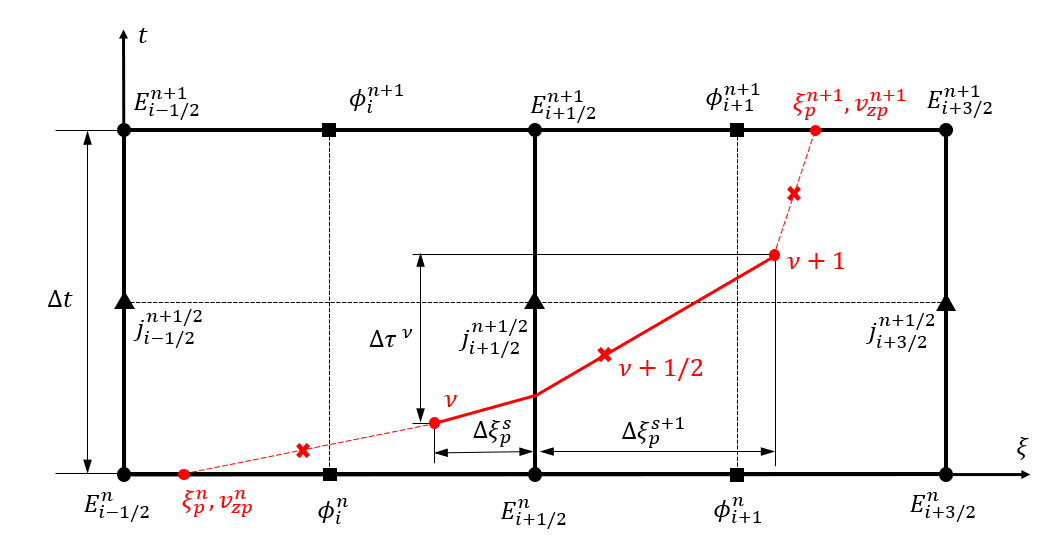}
    \caption{Space-time diagram showing the location of the computed mesh quantities and an example particle trajectory
    (red line), which comprises three suborbits, and the central suborbit (solid red line)  that crosses two different cells. While the velocity $v^{\nu+1/2}_p$ is unique for both segments within this suborbit, their slope in computational space may be different because the map Jacobian $J_\xi$ in general varies from cell to cell.}  \label{fig:trajectory}
\end{figure}

Figure \ref{fig:trajectory} shows an example of a particle's trajectory during and integration time step and the position of the most important field and particle variables in the computational space-time grid. The particles are allowed to cross several cells in each substep $\nu$ instead of stopping at cell faces. This new strategy \cite{chen22} has several advantages. The criteria for the selection of the substep $\tau^\nu_p$ is only based on physical considerations. Free-streaming particles, not subject to strong-field gradients, can be advanced with fewer substeps while a time estimator for $\Delta\tau_p^\nu$ (discussed below) keeps numerical error of the Crank-Nicolson scheme under control.

The electric field (which determines the acceleration) must be scattered to the particles, accounting for all cell crossings during a substep $\Delta \tau^\nu$. In practice, this is done using a segment-averaging approach \cite{chen22}:

\begin{align}\label{eq:scatter}
    E_{\xi,p}^{\nu+1 / 2}=\sum_i E_{\xi,i+1/2}^{n+1/2}\left\langle{{S_1}}\left({\xi}_{i+1/2}-{\xi}_p^{s+1 / 2}\right)\right\rangle_p^\nu,
\end{align}
where ${E}_{\xi,i+1/2}^{n+1/2} = ({E}_{\xi,i+1/2}^{n+1} + {E}_{\xi,i+1/2}^{n})/2$ and the orbit average of the first-order spline function $S_1$ is defined by weighting the distance traveled through each crossed segment ($\Delta \xi^s$) over the total length of the substep ($\Delta \xi^\nu$) \cite{chen22} 
\begin{align}\label{eq:spline}
    \left\langle{{S_1}}\left({\xi}_{i+1/2}-{\xi}_p^{s+1 / 2}\right)\right\rangle_p^\nu=\frac{1}{\Delta \xi_p^\nu} \sum_{s \in\nu} S_1\left(\xi_{i+1/2}-\xi_p^{s+1/2}\right) \Delta \xi_p^s.
\end{align}
Here, the sum is over the collection of segments that make up the substep $\nu$. 

The segment-averaging procedure described above for the electric field in the particle push results in significant efficiency gains with respect to the previous particle subcycling approach \cite{chen11}. The efficiency gain originates in that, on average, the current approach performs a single Picard solve per several cell crossings, instead of a full Picard solve per each cell crossing. The computation of the orbit cell-crossing segments is in practice a very efficient operation once initial and final points are known, and does not significantly burden our algorithm.

\subsubsection{Substep timestep estimator}\label{sec:estimator}

A suitable substep timestep estimator is important to minimize the possibility of particle tunneling across a potential barrier \cite{chen11} (which could happen, for instance, if the timestep is too large for the particle to resolve the scale of variation of the potential barrier). In our implementation, we estimate the timestep to ensure that the truncation error per Crank-Nicolson step is within a specified tolerance. The truncation error $\mathcal{E}_{\Delta \tau}$ of the Crank-Nicolson discretization in Eqs. \eqref{eq:xiCN},\eqref{eq:vCN}  for the position update is given by \cite{chen22}:
\begin{align}
    \mathcal{E}_{\Delta \tau}=\frac{1}{12}\left(\frac{\dd a}{\dd \tau}\right)^v \Delta \tau^3 + O(\Delta\tau^4).
\end{align}
The substep time step can be determined from the desired maximum truncation error in a given substep as \cite{chen22}:
\begin{align}\label{eq:estimator}
    \Delta \tau_p^v=\left(\frac{12 \mathcal{E}_{\Delta \tau}}{|\dd a/ \dd \tau|^v}\right)^{\frac{1}{3}},
\end{align}
where $|\dd a/ \dd \tau|^v$ is to be estimated along the substep as described below. 
In previous studies, the time derivative of the acceleration was computed by pushing the particle using Euler's scheme (constant field) and with a small time step $\delta t = 10^{-8}$, $\left(\frac{d a}{d \tau}\right)^v \simeq\left(a^{\nu+\delta t}-a^{\nu}\right) / \delta t$. This estimate is then used as an initial guess for $\Delta\tau_p^\nu$. However, this approach provides a suitable error estimate only if the particle remains within a cell (where the numerical field is assumed to vary linearly). 
In this study, a given substep may cross several cells, and the estimate of $|\dd a/ \dd \tau|^v$ needs to be adjusted accordingly. Here, we compute it by taking an $L_1$-norm (other norms are possible as well) of the change of acceleration across all cells crossed by the particle as:
\begin{align}
    \left | \frac{\dd a}{\dd \tau} \right |^v = \sum_{s \in \nu} \left | \frac{\dd a}{\dd \tau} \right |_s. \label{eq:segment_acc}
\end{align}
The terms $|\dd a/ \dd \tau|_s$ are computed internally in the particle mover taking the field difference between the initial and final points in the segment.
The substep timestep $\Delta\tau_p^\nu$ \eqref{eq:estimator} is reduced if the prescribed tolerance $\mathcal{E}_{\Delta \tau}$ is not reached. This procedure is outlined in Algorithm~\ref{mover}.

\subsubsection{Orbit integration algorithm}
\label{sec:orbit-int}

The numerical update of the computational coordinate and physical velocity follows from Eqs.\ref{eq:xiCN},\eqref{eq:vCN}:
\begin{align}
&{\xi}_p^{\nu+1}={\xi}_p^\nu+\Delta\tau_p^\nu(v_z/J_{\xi})^{\nu+1 / 2}_p \label{eq:xiupdate}, \\
&v_{z,p}^{\nu+1/2} = v_{z,p}^{\nu} + 0.5\Delta\tau_p^\nu\left(\frac{q_{p}}{m_{p}} \left ( \frac{E_{\xi}}{J_{\xi}} \right )^{\nu+1/2}_p - \Tilde{\mu}_p\frac{B_p^{\nu+1} - B_p^{\nu}}{v_{z,p}^{\nu+1 / 2}\Delta\tau_p^\nu}\right).\label{eq:vupdate}
\end{align} 
After the iteration, the new-time velocity is found as:
\begin{equation}
{v}_{z,p}^{v+1}=2 {v}_{z,p}^{v+1 / 2}-{v}_{z,p}^v.
\end{equation}

The equations are updated until convergence to a very tight relative tolerance, $\mathcal{O}(10^{-13})$,
in $v_{z,p}^{\nu+1/2}$ and $\xi_p^{\nu+1}$. To scatter the field to the position of the particle at mid orbit according to Eq. \ref{eq:scatter}, cell crossings are detected and the distance along cell segments travelled by the particle, $\Delta \xi_p^s$, is computed. Although $B(z)$ is in principle known everywhere in the domain, it is generally faster to interpolate its value at the particle position from stored discrete values at cell faces with a negligible compromise in accuracy, as long as the field is smooth:
\begin{equation}
    B_{p}^{\nu}=\sum_i B_{i+1/2}{S_1}\left({\xi}_{i+1/2}-{\xi}_p^{\nu}\right).
\end{equation}

In previous studies \cite{chac13}, the conversion of the physical velocity for the computation of the new computational position $\xi_p^{\nu+1}$ in Eq. \ref{eq:xiupdate} was performed with $(J_{\xi})^{\nu+1/2}_p$ calculated at the midpoint of the cell in which the particle is located. The new mover takes into account the mesh map across all cells that the particle crosses during a substep. A straight line in the physical space $t-z$ translates into a curved trajectory in the computational space $\xi-t$. However, there still exists a linear relationship between the velocity and the total computational length traveled by the particle, i.e.,  $\Delta\xi_p^\nu\propto(1/(J_{\xi})^{\nu+1/2}_p)v_{z,p}^{\nu+1/2}$ that can be used as the new definition of the mid-orbit Jacobian. In order to find $(J_{\xi})^{\nu+1/2}_p$, we require that the total substep time is equal to the sum of the time the particle spends in each cell, 

\begin{align}
    \Delta \tau_p^\nu = \sum_{s\in \nu}\Delta \tau^s_p.
\end{align}
Using Eq. \eqref{eq:xiCN}, this relationship can be written as:
\begin{align}
    \Delta\xi^{\nu}_p/(v_{z,p}^{\nu+1/2}/(J_{\xi})_p^{\nu+1/2}) = \sum_{s \in \nu} \Delta\xi^{s}_p/(v_{z,p}^{\nu+1/2}/(J_{\xi})_p^{s+1/2}).
\end{align}
Canceling out the velocity on both sides (since it is constant per substep $\nu$), we find a suitable interpolation formula for the Jacobian over a given particle substep:
\begin{align}
    (J_{\xi})_p^{\nu+1/2} = \frac{1
    }{\Delta \xi_p^\nu}\sum_{s\in \nu} (J_{\xi})_p^{s+1/2} \Delta \xi^{s}.
\end{align}
%
Note that the $\Delta \xi^s$ factors were already computed for the scattering of the electric field and are therefore readily available. The Jacobian at the particle position $(J_{\xi})_p^{s+1/2}$ is linearly interpolated from the values at the mesh nodes $J_{\xi,i+1/2}$ 
as:
\begin{align} \label{eq:J_segment}
    (J_{\xi})_p^{s+1/2} =  \sum_i S_1\left(\xi_{i+1/2}-\xi_p^{s+1/2}\right)J_{\xi,i+1/2},
\end{align}
giving:
\begin{align}\label{eq:jacmap}
    (J_{\xi})_p^{\nu+1/2} = \frac{1
    }{\Delta \xi_p^\nu}\sum_{s\in \nu} \sum_i S_1\left(\xi_{i+1/2}-\xi_p^{s+1/2}\right)J_{\xi,i+1/2}\Delta \xi^{s},
\end{align}
which is very similar to Eq. \eqref{eq:spline}, and can therefore be computed along very efficiently. The segment-averaged Jacobian factor in Eq. \eqref{eq:jacmap} is then used in the particle equations of motion, Eqs. \eqref{eq:xiCN} and \eqref{eq:vCN}. A full description of the new particle mover algorithm is provided in Algorithm~\ref{mover}.


\begin{algorithm}[t]
\caption{Particle mover}\label{mover}
    \begin{algorithmic}[1]
    \While {residual > tol}
        \State Compute substep length $\Delta\xi^\nu_p =(v_z/J_{\xi})_p^{\nu+1/2}\Delta\tau^\nu_p$
        \State Determine particle direction and number of crossed cells
        \For {c=1,cells}
            \State Compute segment length $\Delta\xi^s_p$ 
            \State Interpolate electric field and add to segment-averaged value \Comment{Equations \eqref{eq:scatter},\eqref{eq:spline}}
            \State Interpolate $(J_{\xi})_p^{s+1/2}$ and add to segment-averaged Jacobian \Comment{Equations \eqref{eq:J_segment}, \eqref{eq:jacmap}}
            \State Accumulate change in acceleration per segment, $(\dd a/\dd\tau)^s$ \Comment{Equation \eqref{eq:segment_acc}}
        \EndFor
        \State Compute the effective electric field $E_{\xi,p}^{v+1/2}$ \Comment{Equation \eqref{eq:scatter}}
        \State Update velocity $v_p^{\nu+1/2}$ \Comment{Equation \eqref{eq:vupdate}}
        \State Compute the new mid-orbit Jacobian $(J_{\xi})_p^{\nu+1/2}$ \Comment{Equation \eqref{eq:jacmap}}
        \State Compute the residual in position and velocity \Comment{Equations \eqref{eq:xiCN}, \eqref{eq:vCN}}
        \If {$\sum_{s \in \nu} |\dd a/\dd\tau|^s > 12\mathcal{E}_{\Delta \tau}/(\Delta \tau_p^\nu)^3$}
         \State Reduce $\Delta\tau_p^\nu$ 
        \EndIf
    \EndWhile
    
    \end{algorithmic}
\end{algorithm}

\subsection{Field solver}

We now consider the spatial discretization of Eq. \ref{eq:amperecurv}. For convenience and without loss of generality, regular unit cell computational grids will be used (i.e., $\Delta \xi = 1$).
With this choice, $J = \Delta\mathrm{V}_{i}$ is the effective physical cell volume and the standard second-order finite-difference derivative discretization simplifies to
\begin{align}
    \partial_\xi(J j^{\xi})_i\approx (J j^{\xi})_{i+1/2} - (J j^{\xi})_{i-1/2}.
\end{align}
The Laplacian of the scalar potential is discretized using a second-order conservative scheme,
\begin{align}\label{eq:laplacian}
    (\nabla_\xi^2\phi)_i = \partial_\xi \left ( \frac{J^B}{J_{\xi}} \partial_\xi \phi \right )_i \approx \left ( \frac{J^B}{J_{\xi}} \right)_{i+1/2}(\phi_{i+1} - \phi_{i}) - \left ( \frac{J^B}{J_{\xi}} \right)_{i-1/2}(\phi_i - \phi_{i-1}).
\end{align}
The discrete form of Eq. \eqref{eq:amperecurv} reads,
\begin{align}\label{eq:field}
    (\nabla_\xi^2\phi)_i^{n+1}= (\nabla_\xi^2\phi)_i^{n} + \Delta t\left[(J j^{\xi})_{i+1/2} - (J j^{\xi})_{i-1/2}\right]^{n+1/2}.
\end{align}
The solution for the new potential $\phi^{n+1}$ requires the inversion of the Laplacian operator under suitable boundary conditions. From the potential, the covariant component of the electric field is found as: 
\begin{align}
E_{\xi,i+1/2}^n=-(\partial_\xi \phi)_{i+1/2}^n \approx \phi_{i}^n- \phi_{i+1}^n.
\end{align}

The modified current density at the mid timestep is computed as the orbit average along the particle substeps, 
\begin{align}\label{eq:weighting}
(J j^{\xi})^{n+1/2}_{i+1/2}=\frac{1}{\Delta t} \sum_p \sum_{\nu \in n} {(J j^{\xi})}^{\nu+1/2}_{i+1/2, p} \Delta \tau_{p}^\nu,
\end{align}
with:
\begin{align}\label{eq:weighting_p}
    (J j^{\xi})^{v+1/2}_{i+1/2, p}=q_p \left ( \frac{v_z}{J_{\xi}} \right)^{\nu+1 / 2}_p \left\langle{S_1}\left({\xi}_{i+1/2}-{\xi}_p^{s+1 / 2}\right)\right\rangle_p^\nu,
\end{align}
where the same segment-averaged shape function is used as in the scattering of the electric field to the particle positions, Eq. \eqref{eq:scatter}. This finally gives:
\begin{align}
    (J j^{\xi})^{n+1/2}_{i+1/2}=\frac{1}{\Delta t} \sum_p \sum_{\nu \in n} q_p \Delta \xi_p^v \left\langle{S_1}\left({\xi}_{i+1/2}-{\xi}_p^{s+1 / 2}\right)\right\rangle_p^\nu.
\end{align}

\subsection{Conservation properties}\label{sec:consv}
We demonstrate next exact local charge and global energy conservation theorems for the proposed discrete particle representation for the paraxial model.
\subsubsection{Global energy conservation}
From the Crank-Nicolson discretization in Eq. \eqref{eq:vCN},
\begin{align}
    \frac{v_{z,p}^{\nu+1}-v_{z,p}^{\nu}}{\Delta\tau_p^\nu} + \Tilde{\mu}_p\frac{B^{\nu+1} - B^{\nu}}{v_{z,p}^{\nu+1 / 2}\Delta\tau_p^\nu}=\frac{q_{p}}{m_{p}} \left ( \frac{E_{\xi}}{J_{\xi}} \right )_p^{\nu+1/2},
\end{align}
and multiplying by $m_p v_{z,p}^{\nu+1 / 2}$ and recalling that $\mu_p B = m_p v_{\perp,p}^2/2$, we find:
\begin{align}
        \frac{m_p}{2}\left[(v_{z,p}^{\nu+1})^2-(v_{z,p}^{\nu})^2 + (v_{\perp,p}^{\nu+1})^2-(v_{\perp,p}^{\nu})^2\right]=q_{p} \left ( \frac{E_{\xi}}{J_{\xi}} \right )^{\nu+1/2}_p v_{z,p}^{\nu+1 / 2}{\Delta\tau^\nu_p}.
\end{align}
The left-hand side can be easily identified as the variation of the total kinetic energy per particle in a substep $\Delta K^\nu_p$. Summing over all particles and substeps per particle results in:
\begin{align}\label{eq:energy}
\begin{aligned}
    K^{n+1}-K^n &= \sum_p q_p \sum_{\nu \in n}{\Delta\tau^\nu_p} {v}_{z,p}^{\nu+1/2}(E_{\xi}/J_{\xi})_p^{\nu+1/2}\\
    &=\sum_p q_p \sum_{\nu \in n} {\Delta\tau^\nu_p} (v_z/J_\xi)_{p}^{\nu+1/2}\sum_i E_{\xi,i+1/2}^{n+1 / 2} \left\langle\overline{{S}}\left({\xi}_{i+1/2}-{\xi}_p^{s+1 / 2}\right)\right\rangle_p^\nu \\
    &=\sum_i E_{\xi,i+1/2}^{n+1 / 2} (J j^{\xi})^{n+1 / 2}_{i+1/2}  \Delta t \\
    &=-\sum_i (\partial_{\xi} \phi^{n+1 / 2})_{i+1/2} (J j^{\xi})^{n+1 / 2}_{i+1/2}  \Delta t \\
    &=\sum_i \phi_i^{n+1 / 2} (\partial_\xi J j^{\xi})^{n+1 / 2}_{i}  \Delta t \\
    &=\sum_i \phi_i^{n+1 / 2} \epsilon_0 \frac{(\nabla_{\xi}^2 \phi^{n+1})_i-(\nabla_{\xi}^2 \phi^n)_i}{\Delta t}  \Delta t \\
    &=- \frac{\epsilon_0}{2} \sum_i \left [ \phi_i^{n+1} (\nabla_{\xi}^2 \phi^{n+1})_i  -\phi_i^{n} (\nabla_{\xi}^2 \phi^{n})_i \right ]\\
    &=-\frac{\epsilon_0}{2} \sum_i\left[\left ( \frac{J^B}{J_{\xi}} \right)_{i+1/2}\left(J_{\xi}{E_z}^{n+1}\right)_{i+1/2}^2-\left ( \frac{J^B}{J_{\xi}} \right)_{i+1/2}\left(J_{\xi}{E_z}^{n}\right)_{i+1/2}^2\right] \\
    &= -\frac{\epsilon_0}{2} \sum_i\left[\left({E}_{z,i+1/2}^{n+1}\right)^2-\left({E}_{z,i+1/2}^{n}\right)^2\right] J_{i+1/2} \\
    &=-\left(W_E^{n+1}-W_E^n\right),
\end{aligned}
\end{align}
from which total conservation of energy follows.
Above, the second line substitutes the scattering formula for the covariant field, Eq. \eqref{eq:scatter}, the third line commutes the sums and introduces the weighted sum for the current density per each particle and substep $\nu$, Eq. \eqref{eq:weighting}. The fifth equality holds after integration by parts with periodic boundaries (which is the simplest case as it trivially eliminates boundary terms, although the proof can also be extended to reflective boundaries \cite{chen22}), and the discrete field equation, Eq.  \eqref{eq:field}, is introduced in the sixth equality. The seventh equality follows from the self-adjointness of the discrete Laplacian operator. Telescoping the conservative discretization for the Laplacian operator, Eq. \eqref{eq:laplacian}, gives the eighth equality \cite{chac13}, since:
\begin{align}
    \sum_i \phi_i (\nabla_{\xi}^2 \phi)_i & = \sum_i \phi_i \left [ \left ( \frac{J^B}{J_{\xi}} \right)_{i+1/2}(\phi_{i+1} - \phi_{i}) - \left ( \frac{J^B}{J_{\xi}} \right)_{i-1/2}(\phi_i - \phi_{i-1}) \right ]
    \nonumber
    \\
    & = - \sum_i \left ( \frac{J^B}{J_{\xi}} \right)_{i+1/2}(\phi_{i+1} - \phi_{i}) (\phi_{i+1} - \phi_i ) 
    \nonumber
    \\
    & = -\sum_i \left ( \frac{J^B}{J_{\xi}} \right)_{i+1/2} E_{\xi,i+1/2}^2 = -\sum_i \left ( \frac{J^B}{J_{\xi}} \right)_{i+1/2} (J_{\xi} {E_z})^2_{i+1/2} 
    \nonumber
    \\
    & = -\sum_i J_{i+1/2} ({E_z})^2_{i+1/2}.
\end{align}
The ninth equality follows when noting that the term $J=J^BJ_{\xi}$ is the cell volume (since $\Delta \xi = 1$), resulting in the final equality for the change in the total electrostatic energy:

\begin{equation}\label{eq:enedef}
H^n = K^n+W_E^n = K^{n+1} + W_E^{n+1} = H^{n+1}.
\end{equation}

\subsubsection{Local charge conservation}

Next, we show charge conservation per particle per substep $\nu$, 
from which local charge conservation follows. In the continuum, we have the following continuity condition,
\begin{align}
    &\frac{\partial \rho}{\partial t} + \nabla\cdot\mathbf{j} = 0,
\end{align}
which in discrete form and in mapped geometry reads:
\begin{align}\label{eq:chargeconsv}
    \frac{\left(J \rho^{n+1}\right)_i-\left(J \rho^n\right)_i}{\Delta t}+\frac{\left(J j^{\xi}\right)^{n+1/2}_{i+1/2}-\left(J j^{\xi}\right)^{n+1/2}_{i-1/2}}{\Delta \xi}=0.
\end{align}
Pulling the orbit-average sum out, we can write the continuity condition as:
\begin{align}
    \sum_{\nu \in n}\left(\frac{\left(J \rho^{\nu+1}\right)_{i,p}-\left(J \rho^\nu\right)_{i,p}}{\Delta \tau_p^\nu}+\frac{\left(J j^{\xi}\right)^{\nu+1/2}_{i+1/2,p}-\left(J j^{\xi}\right)^{\nu+1/2}_{i-1/2,p}}{\Delta \xi}\right)=0.
\end{align}
At this point, it is sufficient to require charge conservation per substep $\nu$. The charge density term can be decomposed into segment contributions $s$, yielding:
\begin{align}
    \sum_{s \in \nu}\left[\left(J \rho^{s+1}\right)_{i,p}-\left(J\rho^s\right)_{i,p}\right]+\Delta \tau_p^\nu\frac{\left(J j^{\xi}\right)^{\nu+1/2}_{i+1/2,p}-\left(J j^{\xi}\right)^{\nu+1/2}_{i-1/2,p}}{\Delta \xi}=0,
\end{align}
where $s$ and $s+1$ are the start and end points of the segment, respectively. Substituting Eq. \eqref{eq:weighting_p} for the current density and Eq. \eqref{eq:spline} for the segment-averaged spline, we arrive at:
\begin{align}
    \sum_{s \in \nu}\left[\left(J \rho^{s+1}\right)_{i,p}-\left(J \rho^s\right)_{i,p}+q_p\Delta\xi_p^s\frac{S_1\left(\xi_{i+1 / 2}-\xi_p^{s+1 / 2}\right)-S_1\left(\xi_{i-1/2}-\xi_p^{s+1/ 2}\right)}{\Delta \xi}\right]=0,
\end{align}
where we have used that $\Delta \tau_p^\nu ({v_z}/J_{\xi})^{\nu+1 / 2}_p=\Delta\xi^\nu_p$ per Eq. \eqref{eq:xiCN}, which we require to be satisfied per segment. We now introduce the accumulation of the charge density using a second-order B-spline $S_2$ (instead of the first-order spline $S_1$ used for the current) \cite{chen11},
\begin{align}
    (J\rho^{s})_{i, p}=q_p S_2\left({\xi}_{i}-{\xi}_p^{s}\right),
\end{align}
and the final charge-conservation condition reads:
\begin{align}
    \frac{S_2\left(\xi_i-\xi_p^{s+1}\right)-S_2\left(\xi_i-\xi_p^s\right)}{\Delta\xi_p^s} + \frac{S_1\left(\xi_{i+1 / 2}-\xi_p^{s+1 / 2}\right)-S_1\left(\xi_{i-1/2}-\xi_p^{s+1/2}\right)}{\Delta \xi}=0,
\end{align}
which is exactly satisfied by the properties of the B-splines involved \cite{chen11}, proving local charge conservation to machine accuracy.

\section{Magnetic nozzle boundary conditions}\label{sec:bound}
 
At the upstream boundary of a magnetic nozzle, we must prescribe
the distributions of forward particles (i.e., with $v_{z}>0$), and only these. Particles are subsequently allowed to leave the domain through the upstream boundary (reflected particles) or reach infinity downstream (free-streaming particles). In the smooth paraxial expansion, no strong non-neutralities are expected to form.

Numerical simulations  with finite domains must also correctly model the expansion to infinity beyond the end of the domain, 
in particular for the electrons. Since the electrostatic potential $\phi$ continues to decrease beyond the domain boundary to a value at infinity $\phi_\infty$, a fraction of the electrons reaching the boundary of the finite domain will not make it to infinity, and are actually reflected electrons. 
The solution must satisfy a global condition on the electric current leaving the plasma source (e.g., zero current for space plasma thruster applications).
Likewise, BCs must compensate for the different amount of reflected ions and electrons present at the upstream boundary, and resize their input distributions to avoid introducing any spurious non-neutral sheaths into the solution.
Finally, injection must be carried out smoothly to avoid introducing artificial oscillations in the simulation.
 
\subsection{Upstream boundary condition and particle injection}\label{sec:injection}

A homogeneous Dirichlet condition for the potential is imposed at the entrance. Particles are injected through the left boundary with known forward ion and electron distribution functions: 
\begin{align}
f_s^+ = n_s^* \hat f_s^+,
\end{align}
for $s=i,e$, where $\hat f_s^+$ is a normalized distribution function and $n_s^*$ is twice the density of injected particles.
For the verification cases of Section \ref{sec:verification}, the following two-temperature semi-Maxwellian distribution is employed: 
\begin{equation}\label{eq:inj}
    f_s^+\left(z=z_0, v_z>0, \Tilde{\mu}, t\right)=n^*_s\left(\frac{m_s}{2 \pi}\right)^{3/2}(T^*_{\|s})^{-1/2}(T^*_{\perp})^{-1} \exp \left(-\frac{m_s v_z^2}{2  T_{\|s}^*}\right)\exp\left(-\frac{m_s\Tilde{\mu}B_0}{T^*_{\perp s}}\right), \quad s=i, e
\end{equation}
where $T_{\| s}^*$, $T_{\perp s}^*$ are the parallel and perpendicular temperatures of species $s$ and $B_0 = B(z_0)$.

As in other PIC codes \cite{cart00}, 
to model the injection particle flux as accurately as possible,
we sample the component of the velocity normal to the boundary from the flux distribution 
$g_s^+\propto n_s^* v_z \hat f_s^+$ instead of $f_s^+$.
Injected particles must be distributed not only in velocity space but also in time so as to avoid any particle lumping and flux discontinuities caused by a finite integration $\Delta t$. 
An implicit PIC implementation offers a natural procedure for particle injection, even if $\Delta t$ can  be up to two orders of magnitude larger than in classic explicit PIC methods.
A uniformly distributed random integration time step $\Delta t_e\in[0,\Delta t]$ is assigned to each injected particle.
 The injected particles are then pushed as described in Sec. \ref{sec:num} in that time step, 
including the subcycling procedure and their contribution to the moments weighted as for any other particle. 
The injection procedure has a remarkable simplicity compared to other proposed techniques \cite{cart00}
and low computational cost, benefiting from the freedom provided by the implicit mover in the computation of particle's trajectories. 

 
In a MN, some of the injected particles are reflected back in the domain toward the upstream boundary. As the amount of reflected ions and electrons differs in general, 
fixing both $n_i^*$ and $n_e^*$ {\em a priori} can lead to the formation of artificial non-neutral layers at injection, because the integrated density of ions and electrons (traveling forward and backward) may not match. Instead,
$n_e^*$ and/or $n_i^*$ need to be computed as part of the solution to ensure quasineutrality there.
In the present implementation, the injection density of forward ions $n_i^*$ is fixed to produce a prescribed ion current $I_i$, but the injection density of forward electrons $n_e^*$, which  adapt faster to changes in the electrostatic potential due to their larger mobility, is dynamically varied in time according to a simple heuristic proportional control law with gain $G_1$, i.e. $\Delta n_{e}^* = G_1\rho_{i=0}$, where the charge denstiy $\rho_{i=0}$ is weighted at the center of the first cell and accounts for both newly injected and reflected particles. Similar control strategies have been used successfully in previous MN studies~\cite{li19a,sanc18b,meri21a}.


\subsection{Dynamic downstream open boundary}
\label{subsed:dyn_bc}
 
For practical finite domain sizes, part of the potential drop to infiniy occurs beyond the downstream of the domain boundary. This means that a fraction of electrons traversing this boundary are reflected back outside of the domain and eventually return to it. Observe that this does not occur for ions, as the decreasing potential further accelerates them to infinity.

Following a similar approach to that in \cite{li19a}, we sort the electrons that reach the end of the domain ($i=n_z$) according to their kinetic energies. Those with a kinetic energy lower than the remaining potential fall to infinity, $\Delta \phi_\infty = \phi_{i=n_z} - \phi_\infty$, are reflected back, while the rest are removed from the simulation. However, rather than determining $\phi_\infty$ from the current-free condition $j_z=0$ as in \cite{li19a}, we introduce a second control law aimed at offsetting the lack of neutrality in the last cell of the domain $\Delta \phi_\infty = G_2\rho_{i=n_z}$, where $\rho_{i=n_z}$ is the charge density in the last cell of the domain.

The condition $j_z=0$ (or any other given value of the current density in the MN) is instead
implemented through the BC for the electric field. We begin by integrating \eqref{eq:1d_ampere_curv} in $\xi$ to find:
\begin{align}\label{eq:integral_cond}
    J^B(z) \left ( \frac{\partial E_z}{\partial t} + j_z \right ) = C(t),
\end{align}
where we recall that $E_z = E_{\xi}/J_{\xi}$ and $j_z = J_{\xi} j^{\xi}$. The constant $C(t)$ might depend on time but does not depend on position. 
However, particularizing at infinity with $j_z=0$ gives $C(t) = 0$ for all $t$. Taking time-centered finite differences,
\begin{align}
E^{n+1}_{z,n_z+1/2}  = E^{n}_{z,n_z+1/2} -\Delta t j^{n+1/2}_{z,n_z+1/2}.
\end{align}
The new electric field provides an inhomogeneous Neumann boundary condition for the electrostatic potential solver, consistent with the global-current free condition in steady state. 

The proposed approach eliminates the need to prescribe a downstream potential and the subsequent formation of a non-neutral layer as in \cite{sanc18b}, and allows computing the electric field
at the last node self-consistently rather than imposing $E_{i=n_z}=0$ as in \cite{li19a}.



\section{Algorithm verification}\label{sec:verification}

In this section, we present three different verification cases. First, the  conservation properties of the implicit PIC algorithm (\ref{sec:consv}) after the addition of the new segment mover, the paraxial geometry and the magnetic mirror force are demonstrated in an academic periodic magnetic mirror study case. Afterward, we test the fractional injection scheme (\ref{sec:injection}) with a finite (non-periodic) uniform plasma configuration in which the injection plays the role of replenishing the natural outflow of particles through the boundaries. Finally, we put together all pieces and add the new dynamic downstream boundary condition (Sec. \ref{sec:bound}) to study a magnetic nozzle and compare it with the results in the literature. 

\subsection{Periodic magnetic mirror}\label{sec:mirror}

To demonstrate strict conservation properties, we consider a periodic magnetic mirror, where the background magnetic field has a throat (maximum $B$) located at the boundaries of the domain,
\begin{equation}
    B = B_0\left(1 + \frac{R-1}{R+1}\cos\left(\frac{2\pi z}{L}\right)\right),
\end{equation}
where $R=B_{max}/B_{min}=3$ is the mirror ratio and $L=2\pi$ and $z \in [0,L]$.

To begin, we consider singly charged ions and electrons with $m_i/m_e=1$ and equal temperatures. The plasma is homogeneous, with uniform density (equal to unity in normalized units) and isotropic Maxwellian distributions, i.e., Eq. \eqref{eq:inj} with
$T_{s,\|}^* =T_{s,\perp}^*=T_s^*$, and a flat electrostatic potential $\phi$. The initial condition is an exact analytical stable equilibrium.

For the nominal simulation, we consider a non-uniform mesh map with $n_z=64$ and cell size proportional to $B(z)$ (i.e., $J_\xi=B(z)$), and $N_p=1000$ particles per cell per species.
Since the initial condition is a stable equilibrium, any difference with respect to the analytical solution can be attributed to ordinary statistical PIC noise. After running the simulation for $200$ $(\omega_{pe}^*)^{-1}$, we do not observe any secular trend in the simulation, nor instability. The average density in the nominal simulation matches the expected analytical value of unity, and its standard deviation $\sigma[n_e]$ scales
as $1/\sqrt{N_p}$, as expected from the standard Monte Carlo error scaling: for instance, we find $\sigma=2.7\%$ for $N_p=1000$, and $\sigma=1.9\%$ for $N_p=2000$, i.e., a decrease of $1/\sqrt{2}$. For this example, we do not observe any noticeable dependence of the noise level with $\Delta t$ when varied from 1 to 10 $(\omega_{pe}^*)^{-1}$.

\begin{figure}[h]
\centering
\includegraphics[width=\textwidth]{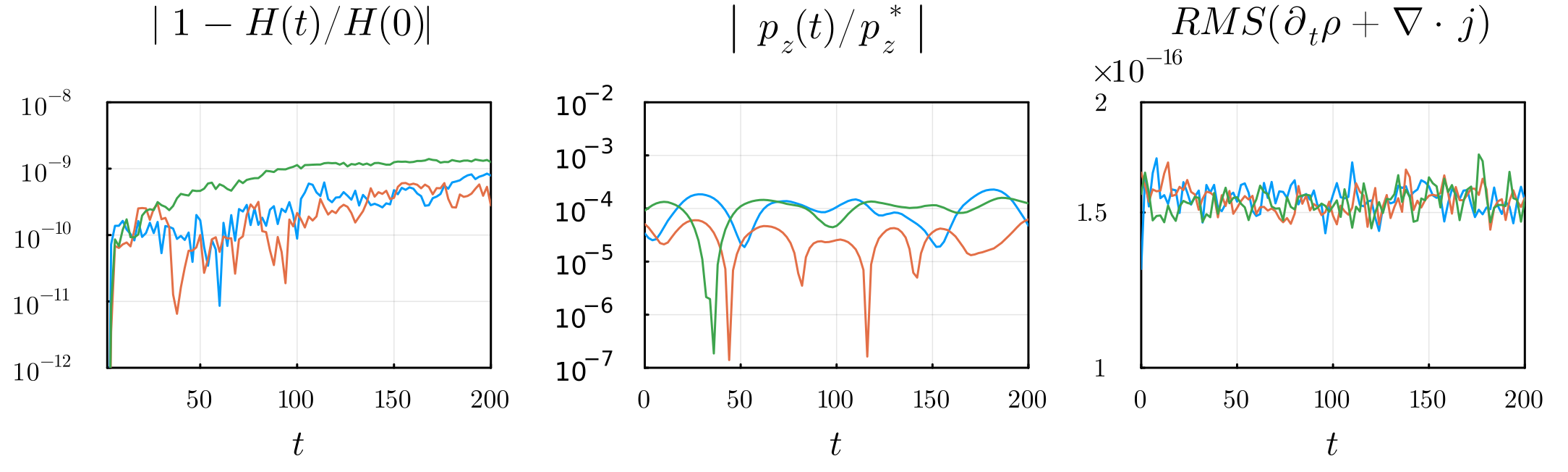}
\caption{Self-consistent periodic magnetic mirror solution for different parallel-to-perpendicular temperature ratios: 
$T_\perp^*=T_\|^*$ (green), $T_\perp^*=2T_\|^*$ (orange), and $T_\perp^*=T_\|^*/2$ (blue). From left to right: conservation of global energy, total momentum, and root mean square of the local charge residual.}\label{fig:mirror_consv}
\end{figure}

Figure \ref{fig:mirror_consv} shows the evolution of the relative error in the conservation of global energy, total axial momentum, and spatial root mean square of the residual in the charge conservation equation. The instantaneous total energy of the system is the sum of the kinetic energy of each particle and the electric field energy as defined in Eq. \eqref{eq:energy}, while the total axial momentum of the plasma is given by:
\begin{equation}
    p_z =  \sum_s m_s\sum_p w_pv_{z,p,s}.
\end{equation}
The relative error in energy conservation remains below $10^{-9}$, which is considerably lower than the upper bound imposed by the relative tolerance of the non-linear solver ($10^{-6}$). In the presence of an externally applied magnetic field, the momentum of individual particles is not generally conserved (indeed, this is the contactless thrust generation mechanism in a MN). Nevertheless, the symmetry of this periodic configuration results in the cancellation of the axial momentum particle contributions, so that $p_z(t)=0\;\forall t$. The error in total axial momentum conservation, normalized with $p_z^*=\sum_s\sqrt{T_s^*} m_s\sum w_p$, is kept acceptably low and does not present secular trends. To that end, we must mention the role of the substep time step estimator (Sec. \ref{sec:estimator}) in keeping the truncation error of the Crank Nicolson mover within the defined tolerance ($\mathcal{E}_{\Delta \tau}=10^{-3}$).  Finally, the root mean square (RMS) of the residual of the local charge conservation equation (computed as indicated in Sec. \ref{sec:consv}) computed over all cells shows errors at machine accuracy ($\sim10^{-16}$).

\begin{figure}[ht]
\centering
\includegraphics[width=0.7\textwidth]{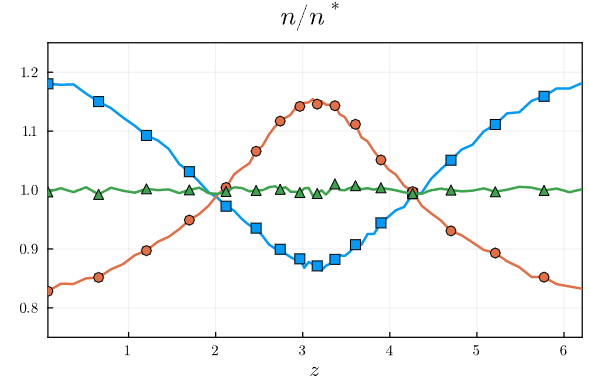}
\caption{Steady-state plasma density over initial plasma density for different parallel-to-perpendicular temperature ratios: $T_\perp^*=T_\|^*$ (\mytriang),
$T_\perp^*=2T_\|^*$ (\mycirc),  and $T_\perp^*=T_\|^*/2$ (\mysquare).}\label{fig:mirror}
\end{figure}

Adding temperature anisotropy to the initial condition results an initial transient, eventually leading to nontrivial density profiles within the mirror, as shown in Figure \ref{fig:mirror}. Considering $T_\perp^*/T_\|^*>1$ leads to a density profile that peaks at the minimum of the magnetic field $B$, as more particles are trapped by the magnetic mirror. Conversely, $T_\perp^*/T_\|^*<1 $ results in the highest density at the magnetic throat, as the number of free or passing particles increases. In all cases, as evidenced in Figure \ref{fig:mirror_consv}, conservation of the three relevant quantities behave similarly to the $T_\perp^*/T_\|^*=1$  case. 
  
\subsection{Particle injection in a uniform plasma}\label{sec:uniformplasma}

We verify next the particle injection BC 
without the generation of artificial electrostatic oscillations or particle lumping. We consider  a uniform Maxwellian plasma at rest in a uniform $B$ field. 
To compensate for particles leaving the domain, the injection BC is applied at the two ends of the domain, 
where a semi-Maxwellian distribution with the same parameters as the initial distribution is prescribed for both ions $i$ and electrons $e$. As before, we set isotropic and equal temperatures for both species and select a mass ratio $m_i=m_e=1$. The same non-uniform mesh as in the mirror case is employed ($n_z=64$), with the same number of particles per cell and species, $N_p=1000$.  Periodic BCs are applied to the electrostatic potential.
The initial condition is again a stable exact equilibrium.

\begin{figure}[ht]
    \centering   \includegraphics[width=0.5\textwidth]{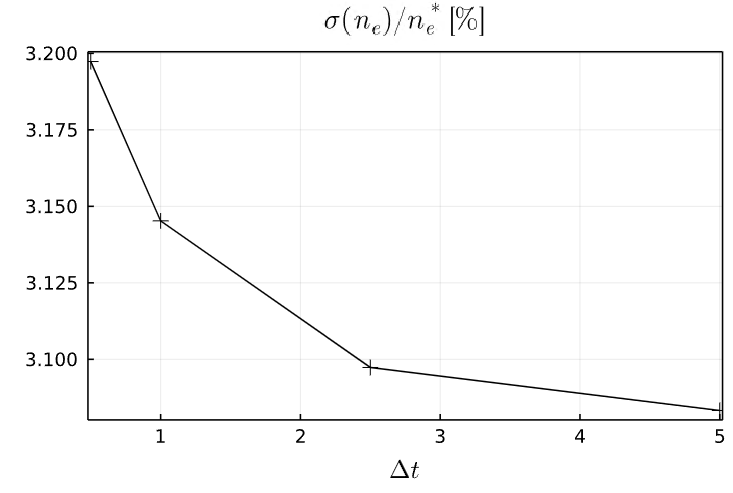}
    \caption{Standard deviation of the electron density across all cells and timesteps with respect to the analytical solution ($n_i(z) =n_e(z) = n_e^* \;\forall\;z$) as a function of the implicit timestep size $\Delta t$.}
    \label{fig:inj_sigma}
\end{figure}

Plasma uniformity is recovered, except for particle noise. The standard deviation $\sigma(n_e)$  of the plasma density with respect to the uniform value $n_e^*$ across all time steps and cells is shown in Figure \ref{fig:inj_sigma}. The noise level is again consistent with Monte Carlo estimates ($1/\sqrt{N_p}\sim3\%$), and shown to decrease slightly with increasing $\Delta t$ in the range 1 - 5 $(\omega_e^*)^{-1}$. We hypothesize that this behavior is due to the sampling multiplication effect of the orbit averaging of the current density moment scatter in the implicit time-marching scheme, but at any rate its impact is small.

\subsection{Magnetic nozzle modeling}\label{sec:nozzle}
 
We simulate next the plasma expansion in a propulsive magnetic nozzle. We aim to verify
the interoperation of all elements of the algorithm, and in particular,
the dynamic downstream boundary condition.
We compare our simulation results with the nominal MN simulation of \cite{sanc18b}. That study presents a solution of the fully magnetized plasma flow in a paraxial convergent-divergent magnetic field,  using 
a time-explicit semi-Lagrangian scheme. Consequently, it provides a suitable benchmark for the comparison and verification of our algorithm.

The magnetic topology in \cite{sanc18b} is generated by 
a current loop of radius $r_L$ placed at $z=0$,
\begin{equation}
    \boldsymbol{B}(z)=B_0\frac{r_L^3}{\left(r_L^2+z^2\right)^{3 / 2}}\boldsymbol{1}_z.
\end{equation}
The domain
extends from $z=-0.5 r_L$ to $z=16 r_L$, where it is terminated by a dielectric wall.
Distribution functions for forward electrons and ions are prescribed at the domain entrance as in equation \eqref{eq:inj},
with $T_{\|e}^*=T_{\perp e}^*=T_{\|i}^*=T_{\perp i}^*\equiv T^*$.
To eliminate non-neutralities at injection, $n^*_i$ is varied depending on the number of reflected electrons, while $n_e^*$ is fixed.
Both boundaries are free-loss surfaces for outgoing particles. A reduced mass ratio $m_i/m_e=100$ is used.  
The normalized Debye length based on $n_e^*$ and $T^*$ is $\lambda_D^*/r_L = 0.02$, which is quite large in relation to devices of practical interest. In Ref. \cite{sanc18b},
Dirichlet boundary conditions are applied on the electrostatic potential at the throat $\phi_t(z=0)=0$ and at the downstream boundary, where it is adjusted iteratively after convergence to steady state to achieve a current-free expansion. Upon convergence, the reference finds $I_i=-I_e\approx0.074 en_e^*\sqrt{T_e^*/m_e}$. 
A variable mesh with 1501 nodes in $z$, 77 in $v_\parallel$ and 101 in $\mu$; and a time step of 0.03 $(\omega_{pe})^{-1}$ are used.
 
\subsubsection{MN simulation setup}

\begin{figure}[h]
    \centering
    \includegraphics[width=0.75\textwidth]{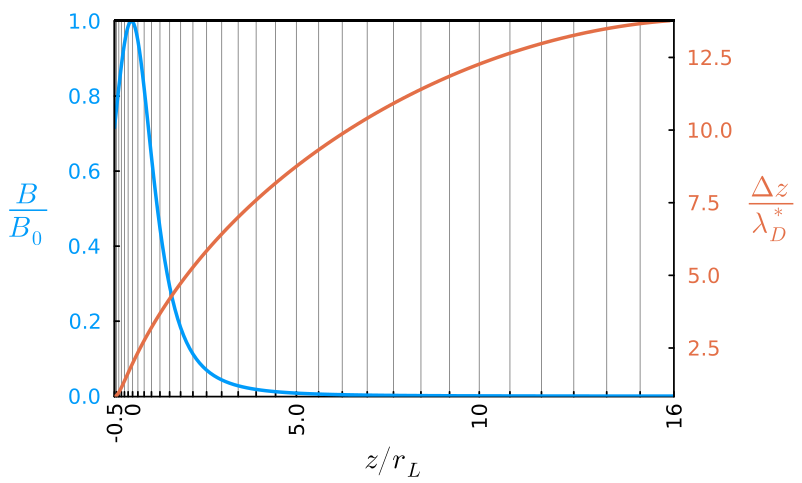}
    \caption{Simulation domain for the MN verification case: Magnetic field strength mesh and cell spacing for the nominal case.}
    \label{fig:domain}
\end{figure}

In what follows, we discuss two nominal simulation cases: 
\begin{enumerate}
\item[A] \textbf{Dielectric downstream wall}: 
This simulation reproduces the physical setup of the simulation in \cite{sanc18b}, described above. 
To model the absorbing dielectric wall, the dynamic downstream electron reflection control is turned off, and all electrons reaching the end of the domain are removed, while the current-free condition on the electric field of section \ref{subsed:dyn_bc} is maintained. As a consequence, a non-neutral layer forms downstream as in \cite{sanc18b}.
The comparison with \cite{sanc18b} and \cite{meri21a} allows us to verify all elements of the implicit algorithm, except for the electron reflection control.
 
\item[B] \textbf{Expansion to infinity}: By activating the dynamic downstream electron reflection control, we simulate the expansion of the plasma to infinity while avoiding the formation of any non-neutral layers. 
\end{enumerate}
We will also consider refined versions of these (which we will identify as AF, BF), in which all numerical parameters are refined to provide an appropriate baseline for error estimation (i.e., smaller time step, larger mesh size, larger number of particles, and tighter tolerances). The specific numerical parameters used in these simulations are reported in Table \ref{table:simparams}. The only setup difference between A and B is the activation of the downstream electron reflection control ($G_{2,B} \neq 0$), allowing the simulation of an expansion to infinity.
The spatial meshes used have variable cell
size to exploit the variable gradient length expected in expansion; the nominal mesh for A and B is shown in Figure \ref{fig:domain}. Simulations AF and BF are intented as a reference to test the convergence of our results, discussed in Section \ref{sec:conver}. 
\setlength{\arrayrulewidth}{0.8pt}
\begin{table}[ht!]
\centering
\caption{Nominal Simulation Parameters. The minimum of $N_p$ occurs at the entrance (we observe about a four-fold increase in $N_p$ at the exit for our non-uniform mesh, Figure \ref{fig:domain}).}
\begin{tabular}{ccc}
\toprule[2pt]
Numerical Parameter   & Nominal (A and B) & Fine (AF and BF) \\
\midrule[1.5pt]
\setlength{\arrayrulewidth}{0.4pt}
Number of cells $n_z$   & 128 & 256 \\
Cell length at entrance & 1.5  $\lambda_D^*$ & 0.75 $\lambda_D^*$\\
Cell length at exit     & 12.5  $\lambda_D^*$ & 6.25 $\lambda_D^*$\\
Integration time-step $\Delta t$  &  5.0  $(\omega_{pe}^*)^{-1}$ & 2.5 $(\omega_{pe}^*)^{-1}$ \\
Simulation time  & 12500 $(\omega_{pe}^*)^{-1}$ & 12500 $(\omega_{pe}^*)^{-1}$\\
Minimum particles/cell steady state $N_{pe}\approx N_{pi}$ & $\sim1000$ & $\sim2000$  \\
Total particles at steady state &  $\sim6\cdot10^5$ & $\sim2.4\cdot10^6$ \\
JFNK solver tolerance & $10^{-4}$ &  $10^{-6}$\\
Maximum substep estimator truncation error $\mathcal{E}_{\Delta \tau}$ & $10^{-3}$ & $10^{-5}$ \\
Injection quasineutrality control gain $G_1$ & 0.5 & 0.125 \\
Downstream electron reflection control gain $G_2$ & 0.0 (A) 0.025 (B) & 0.0 (AF) 0.0125 (BF) \\
\bottomrule[2pt]
\end{tabular}
\label{table:simparams}
\end{table}
\subsubsection{MN transient plume expansion}

\begin{figure}[ht]
\centering
\includegraphics[width=\textwidth]{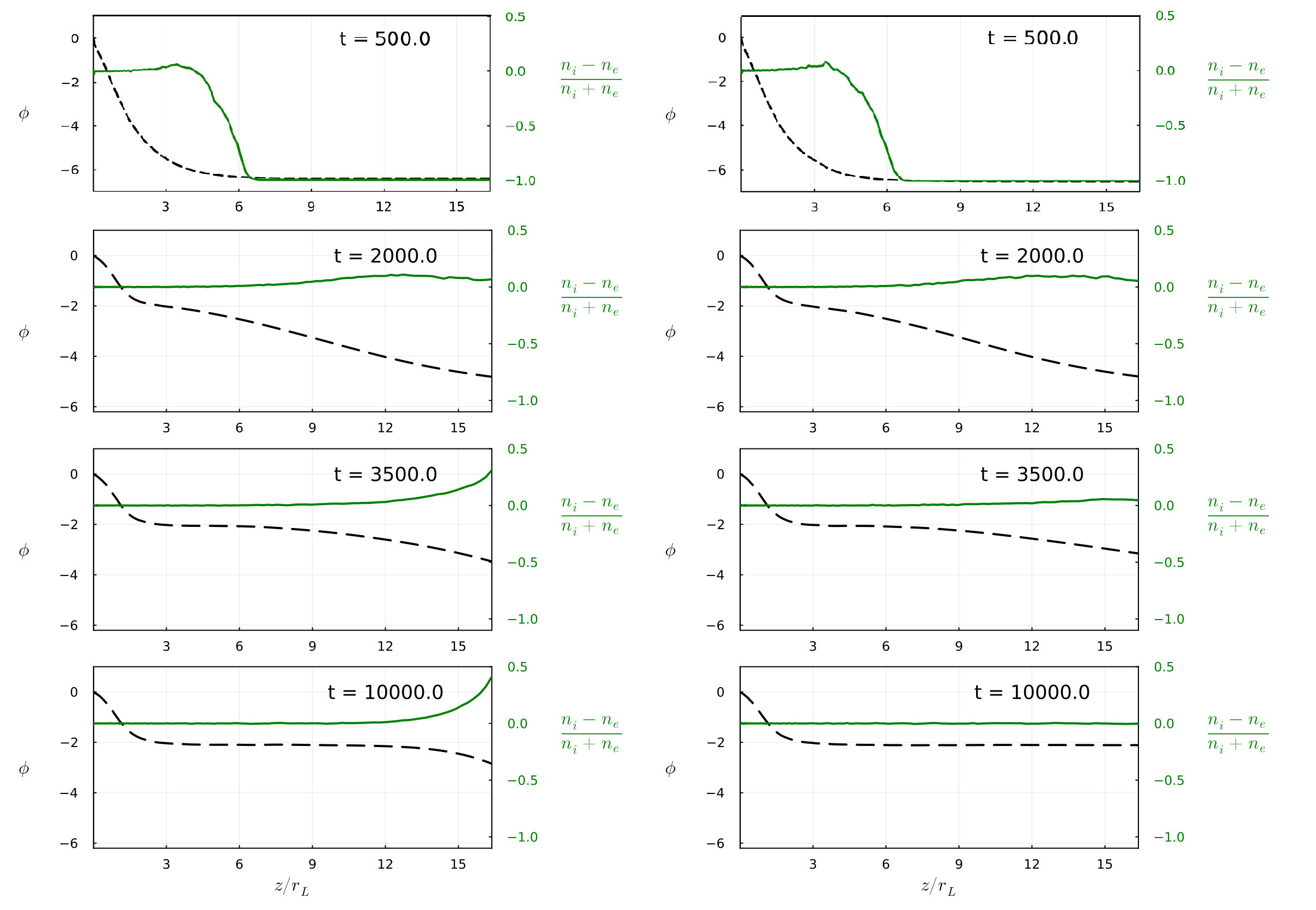}
\caption{Evolution of the potential for simulation cases A (left column) and B (right column).}
\label{fig:evolution}
\end{figure}

Figure \ref{fig:evolution} shows four snapshots of the evolution of the electrostatic potential $\phi$ and the charge imbalance $(n_i - n_e)/(n_i+n_e)$ in the domain for simulations A and B.
In the initial stages [$t=500\;(\omega_{pe}^*)^{-1}$], the lighter electrons rush downstream ahead of the heavier ions, creating a  expansion front with negative charge followed by positive charge. 
During the transient plume expansion, electrons bouncing off the moving expansion front, with its sharp potential drop, lose mechanical energy to the time-varying field. This promotes the formation of a trapped population of electrons \cite{mart15a,sanc18b,meri21a}, discussed later in this study.

Both simulations follow a similar response until the electron/ion expansion front reaches the end of the domain.
Around $t=2000\;(\omega_{pe}^*)^{-1}$, the electron front has left the domain, while the ion front is still traveling downstream, resulting in significant charge separation downstream. 
Thereafter, the two simulations diverge. Around $t=3500\;(\omega_{pe}^*)^{-1}$, a non-neutral thick sheath has begun to form in simulation A, whereas non-neutrality has started to vanish in simulation B owing to the dynamic downstream dynamic electron control, which gradually adjusts the value of the electrostatic potential downstream to promote quasineutrality in the last mesh cell.
Finally, a steady-state sheath forms in simulation A, and a smooth, quasineutral steady-state solution develops in simulation B, as illustrated for $t=10000\;(\omega_{pe}^*)^{-1}$.

\begin{figure}[h]
    \centering    \includegraphics[width=0.8\textwidth]{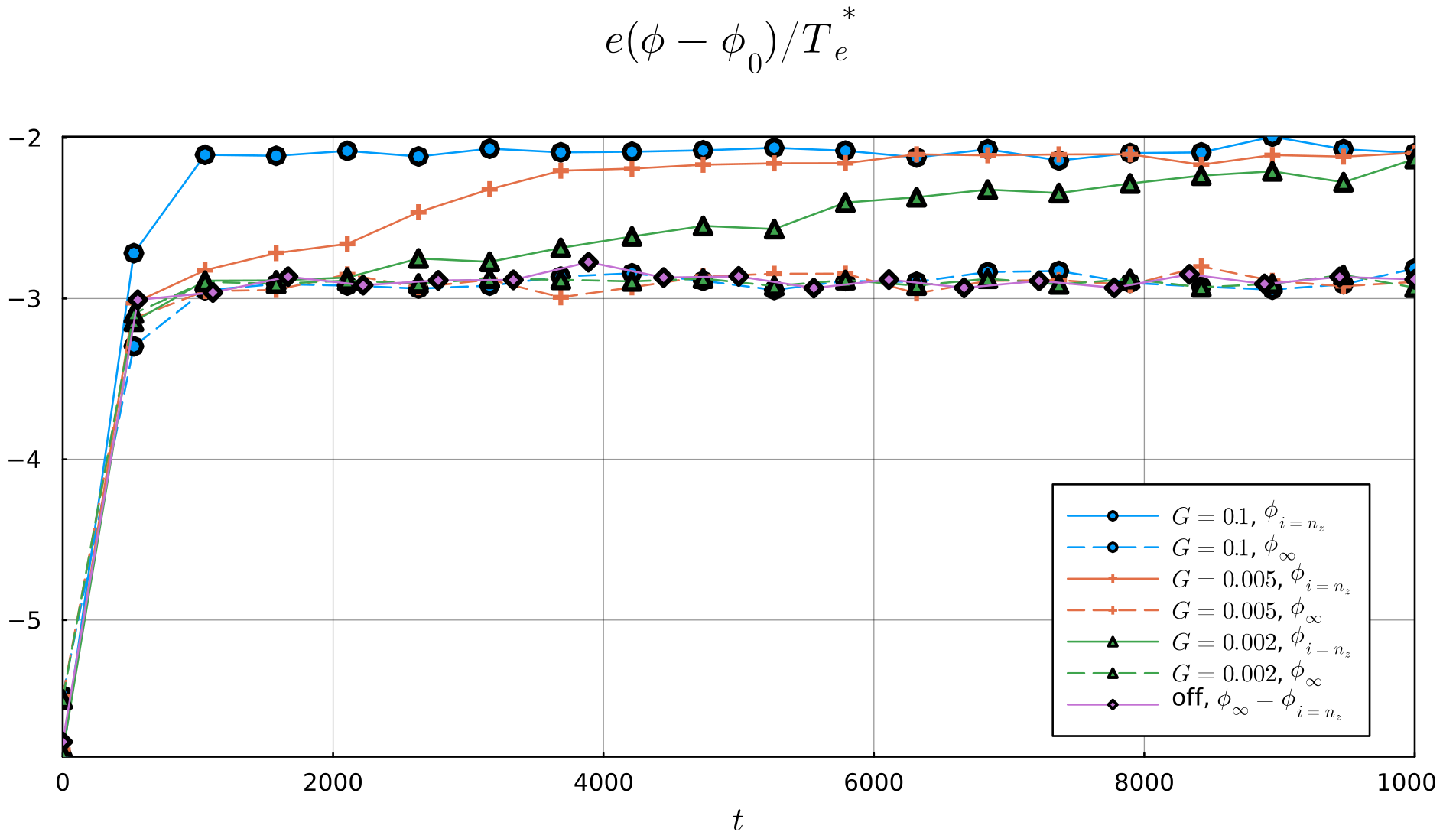}
    \caption{Potential drop at infinity  $e(\phi_\infty-\phi_0)/T_e^*$ (dashed) and at the last cell of the domain $e(\phi_{i=n_z}-\phi_0)/T_e^*$ (solid) for different electron reflection control gains.}
    \label{fig:sensitivity}
\end{figure}

For simulation B, the duration of the transition to quasineutrality is governed by the reflection control parameters 
(downstream control gain and initial critical energy).
Figure \ref{fig:sensitivity} shows the evolution of the potential at the last node of the mesh, $\phi_{i=n_z}$, and at infinity, $\phi_\infty$,
for different control gains $G_2$.
As can be observed, 
once particles start to reach the downstream boundary,
$\phi_\infty$ quickly stabilizes around its asymptotic value, which is independent of the gain $G_2$ chosen.
This independence is expected, as  $\phi_\infty$  controls the net electron current in the MN, and therefore the global current-free condition determines $\phi_\infty$ as a function of ion current and the thermal electron speed at injection \cite{meri21a}. 
The potential at the last node, $\phi_{i=n_z}$, 
also tends to an asymptotic value that is independent of $G_2$, but
its time evolution is sensitive to this parameter, with lower values of $G_2$ corresponding to longer transients.
For this simulation case, gains around $G_2\simeq 0.1$ enable reaching the steady-state value of $\phi_{i=n_z}$ roughly at the same time as the steady-state value $\phi_\infty$, and therefore allow for a fast convergence to the final solution.
The maximum value of $G_2$ is limited by the appearance of instabilities in the solution, which occur when the characteristic control frequency becomes comparable to the other frequencies of the problem.  

Varying the normalized Debye length $\lambda_D^*/r_L$ while keeping all other parameters unchanged (not shown) proves that $\lambda_D^*/r_L$ influences the size of the non-neutral layer downstream in simulation A,
but plays a negligible role in simulation B, because the full expansion is quasineutral in the latter. 

\subsubsection{MN steady-state solution}

\begin{figure}[ht!]
    \centering
    \includegraphics[width=\textwidth]{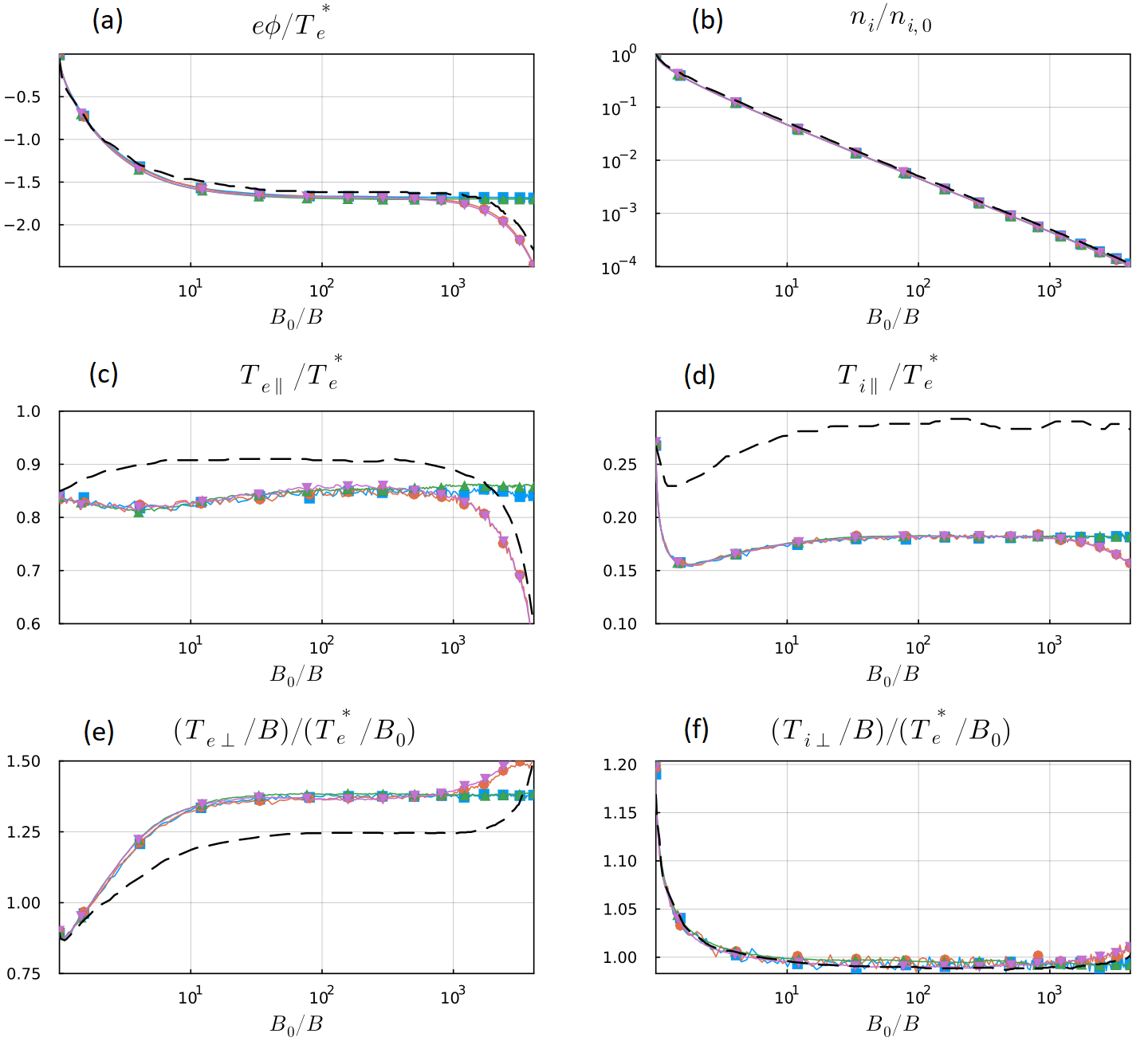}
    \caption{Steady-state potential (a) ion density (b), electron (c) and ion (d) parallel temperatures, and electron (e) and ion (f) perpendicular temperatures for nominal simulations A (\mycirc), B (\mysquare) and refined simulations \textit{AF} (\myitriang) and \textit{BF} (\mytriang). The black dashed lines correspond to the results of Sanchez et al. \cite{sanc18b}. 
    }
    \label{fig:steady}
\end{figure}

Figure \ref{fig:steady}(a) depicts the steady state results of simulations A and B.
The results are time-averaged  using a moving mean with window $\sim500\;(\omega_{pe}^*)^{-1}$ for smoother results.
The decreasing electrostatic potential $\phi$ reflects electrons upstream and accelerates ions downstream, helping to convert the thermal energy of electrons into the directed kinetic energy of ions.
The first part of the expansion is essentially identical for simulations A and B: most of the potential drop occurs here, and
between the throat $z=0$ (where e($\phi_0 - \phi_t)/T_e^*\approx0.4$) and $z=2r_L$ we find the region of maximum acceleration, with an almost constant electric field. After this point, the potential transitions to an essentially flat profile around $z=3r_L$, with $e(\phi_0-\phi)/T_e^* \approx 2.1$.
The differences between simulations A and B arise in the last part of the expansion, where simulation A displays the wide sheath before the dielectric wall,
whereas simulation B continues its slow decrease to infinity. 
It should be noted that the wall potential in simulation A  and the potential at infinity for simulation B are virtually identical ($\approx2.9$ e$/T_e^*$).
 
The ion density $n_i$ is shown in Fig. \ref{fig:steady}(b) from the throat to the end of the domain. After a region of rapid acceleration, driven by the strong electric field near the throat, the ion velocity approaches an asymptotic value and the ion density becomes nearly proportional to $B$, in agreement with the one-dimensional steady-state continuity equation $n_iu_i/B=const$. The ion density exhibits the same trend in the two simulations A and B, with negligible differences.
The electron density $n_e$, not shown here, follows closely the ion density everywhere except in the downstream sheath in simulation A. 

Figures \ref{fig:steady}(c) and (d) show $T_{e\|}$ and $T_{i\|}$, while panels (e) and (f) present the ratios $T_{e\perp}/B$ and $T_{i\perp}/B$.
This presentation choice makes manifest that in an expansion to infinity, the parallel temperatures asymptote to a constant, while the perpendicular temperatures vanish as $O(B)$  \cite{ramo18a}.
Indeed, as it can be observed, the plotted quantities are quite uniform in most of the domain, except in the initial part of the expansion where rapid changes occur.
Once again, simulations A and B display a near-identical behavior except in the last part of the domain, where simulation A is affected by the formation of the thick sheath downstream.
 
\begin{figure}[h]
\begin{subfigure}{.5\textwidth}
    \centering
    \includegraphics[width=\textwidth]{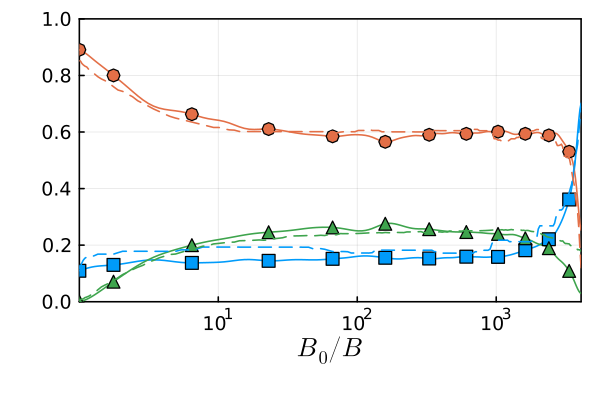}
    \caption{Simulation A}
    \label{fig:tOff}
\end{subfigure}
\begin{subfigure}{.5\textwidth}
    \centering
    \includegraphics[width=\textwidth]{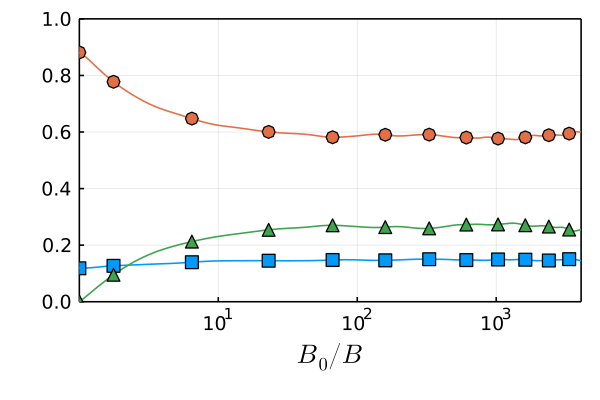}
    \caption{Simulation B}
    \label{fig:tOn}
\end{subfigure}
\caption{Fraction of free $n_{ef}/n_e$ (\mysquare), reflected  $n_{er}/n_e$ (\mycirc) and trapped  $n_{et}/n_e$ (\mytriang) in simulations A (left) and B (right). The dashed lines   correspond to the results of Sanchez et al. \cite{sanc18b}.}
\label{fig:trapped}
\end{figure}

Figure \ref{fig:trapped} shows the steady-state densities of free, reflected, and trapped electrons as defined in \cite{mart15a} and in Section \ref{sec:mod}, for simulations A and B.
As the population of trapped electrons is disconnected from the plasma source,
its distribution depends on the transient plume expansion, wherein some electrons bouncing off the expansion front lose enough mechanical energy to become trapped.
Our investigations shows that the fraction of trapped electrons is essentially independent of the control gain $G_2$ in simulation B, which determines the rate at which the quasineutral steady-state solution is reached. 
In the initial part of the expansion, reflected electrons account for roughly 90\% of the total, while the number of trapped electrons is zero. The fractions then stabilize around 60\% and 25\%, respectively.
Free electrons constitute roughly 10--15\% of the total throughout the expansion.
Simulations A and B show a nearly identical response in terms of electron population fractions, except ---once again--- downstream, where the thick sheath in simulation A reduces the fraction of reflected and trapped electrons to zero at the end of the domain, and all electrons become free.
Arguably, this same transition to free electrons must also occur in simulation B in the expansion to infinity, albeit the scale at which this takes place is too large for practical interest.

\subsubsection{MN solution verification}\label{sec:conver}

We verify our MN solution in two ways. Firstly, we compare the steady state A and B simulation results against the refined simulations AF and BF (Table \ref{table:simparams}) in figure \ref{fig:steady}.
It is visually apparent that the differences are small, with  $T_{\parallel e}$ showing the largest differences, below $3 \%$ in the central part of the expansion for A and the downstream boundary for B, where electrons are reflected.
We therefore conclude that the A and B simulation results are converged. 

Secondly, we compare our simulation A results against the recent results documented in Ref. 
\cite{sanc18b}. This comparison is also shown in figure \ref{fig:steady}.
The normalized electrostatic potential $e\phi/T_e^*$ in figure
\ref{fig:steady}(a)
demonstrates excellent agreement upstream. However, a difference  of $\approx 0.25$ $T_e^*/e$ soon develops that continues all the way to the wall.
This difference is noteworthy, because the final value of the potential governs the current of free electrons, which must coincide with the ion current as dictated by the current-free condition in the MN, as discussed above. This condition is well satisfied in our conservative simulations.

Relatedly, as shown in figure \ref{fig:trapped}, the fractions of free, reflected, and trapped electrons, while showing overall good agreement with those in Ref. \cite{sanc18b}, display small but significant differences in the fraction of free electrons, which is consistently larger in the data from \cite{sanc18b}.
This larger fraction is
consistent with a smaller total potential drop in that study. 

These discrepancies in $\phi$ and the mismatch in the fraction of free electrons explain the  the rest of plasma variables shown in figure \ref{fig:steady}:
While the ion density coincides well with the verification data,
the latter is consistently about $5\%$ higher downstream.  
The comparison of electron and ion temperatures shows that, although the qualitative trends and behaviors remain similar, there are apparent differences in the electron perpendicular and parallel temperatures and in the ion parallel temperature. Plasma density, but especially electron and ion temperatures, are particularly sensitive to the composition of the electron population \cite{meri21a}.
 
Further comparison against the work of Merino et al. \cite{meri18a,meri21a} 
shows that the total potential drop in our simulations is within $0.05$ $T_e^*/e$ of those steady-state results, even if
they do not simulate the region before the magnetic throat.
The smaller discrepancy in this key quantity, together with the conservation properties of the algorithm and the good convergence of our simulations upon numerical parameter variation, supports the verification of our simulations.

\subsubsection{Efficiency and performance}\label{sec:cost}

\begin{figure}[ht!]
    \centering
    \includegraphics[width=\textwidth]{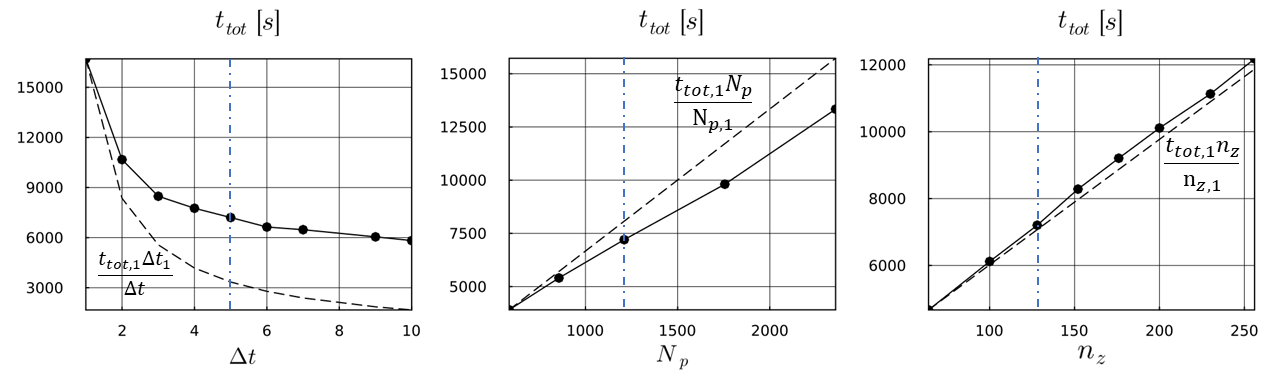}
    \caption{Total CPU wall time as a function of $\Delta t$ (left), the number of particles per cell at steady state $N_p$ (center) and  number of cells in the domain with a constant number of particles per cell $n_z$(right). Reference scaling laws are shown with dashed lines in each plot.}
    \label{fig:times}
\end{figure}

Several numerical parameters
(timestep size, number of particle per cell, and number of cells) 
have been varied in the nozzle simulation to study their impact on the total CPU wall-clock time. The results are shown in Figure \ref{fig:times}. Wall-clock times were obtained with our Julia code, using 20 threads on an Intel Xeon (R) 4316 computer. 

In terms of the implicit timestep size (figure \ref{fig:times}-left), performance is affected by competing effects. On the one hand, an increase in $\Delta t$ may result in an increase in the number of suborbits per particle; it may also result in an increase of the number of function evaluations $N_{FE}$ (each requiring a full particle push) as the number of JFNK iterations increases for convergence. 
On the other hand, there is an inverse proportional relationship between $\Delta t$ and the total number of steps. The balance between these effects determines the wall-clock time speedup as a function of $\Delta t$. As in earlier implicit PIC studies \cite{chen14}, we observe computational advantage for moderate implicit timesteps, but eventual saturation when $\Delta t$ is increased beyond $\Delta t > 10 (\omega_{pe})^{-1}$.

In terms of the total number of particles (or the number of particles per cell $N_p$ when the number of cells is kept constant; figure \ref{fig:times}-center), the CPU wall-clock time scales sublinearly with the number of particles for moderate number of particles, but eventually recovers linear scaling, as expected. The transient sublinear scaling is attributed to the cost contribution of mesh-related operations, which are independent of the number of particles.

The most remarkable result is the scaling of the wall-clock time with the number of cells, $n_z$ (figure \ref{fig:times}-right). In earlier implicit PIC studies employing the subcycling particle push \cite{chen14}, it was predicted and observed that the computational complexity scaled as $n_z^2$. At the root of this behavior is the particle subcycling, and in particular the need for particles to stop at cell faces for charge conservation. The particle mover proposed in this study removes this requirement, decoupling the cost of particle motion from the underlying mesh. As a result, figure \ref{fig:times}-right demonstrates an almost perfect linear scaling of the wall-clock time with $n_z$ while keeping the number of particles per cell constant. Also important for this scaling is the fact that JFNK performance is largely independent of $n_z$, a consequence of the inversion of the Laplacian operator in the residual (Eq. \ref{eq:ampere-inv}), which adequately removes numerical stiffness stemming from this operator.

We estimate next the wall-clock speedup of the implicit PIC algorithm vs. explicit PIC. Speedup may originate from two sources: longer timesteps and decreased mesh resolution. We first consider the timestep. The nominal simulation 
(indicated by the blue vertical line in the graphs of figure \ref{fig:times}) runs in about 2 hours and takes about $N_{FE} \sim 14$ function evaluations per timestep, largely independent of the mesh refinement (this number can be improved further by better preconditioning \cite{chen14}, which we will consider in future work). The explicit PIC timestep is typically $\Delta t = 0.1(\omega_{pe})^{-1}$. For this timestep, the cost per particle push decreases due to a reduction in the number of suborbits per particle (and segments per suborbit); we consider here a reduction of a factor of 2. From there results an implicit-to-explicit speedup from the timestep of $5/0.1 \times 1/14 \times 1/2  \sim 2$. This seems to be a conservative estimate, as figure \ref{fig:times}-left already shows a speedup $\sim 2$ from $\Delta t = 1$ to $\Delta t = 5$.
Regarding mesh resolution, explicit PIC codes typically employ a uniform mesh and require resolving the Debye length to avoid finite-grid instabilities, with $\Delta z < \pi\lambda_D$ \cite{BIRD91}. Taking the Debye length at the MN entrance 
as representative, a uniform mesh of $n_z\approx800$ nodes would be needed for the same problem. Assuming the same number of particles per cell was used and taking a linear growth in computational time with the number of cells (figure \ref{fig:times}-right), there results a performance gain from our adaptive mesh of $800/128\sim6$. This finally leads to an estimated implicit vs. explicit speedup of $\mathcal{O}(10)$. 
 
Another useful datapoint is the wall-clock time comparison vs the semi-Langrangian Vlasov code in Ref. \cite{sanc18b}. That study employed 1501 cells in $z$ and took approximately 2.5 days on a comparable machine with threading \cite{sanchez-pc}. This gives a speedup of $\mathcal{O}(30)$ per simulation for comparable or superior accuracy.  

\section{Conclusions and future work}\label{sec:concl}

A novel implicit particle-in-cell paraxial model has been proposed, targeting the simulation of magnetic nozzles in electric propulsion systems. The new model has demonstrated exact conservation properties in closed systems,
and the ability to employ, stably, much longer timesteps than inverse plasma frequencies and mesh spacings much larger than the Debye length. As a result, the model employs  fewer mesh cells, with a significant reduction in the total number of macroparticles required for the simulation and a corresponding decrease in computational time (which scales linearly with the number of total particles). The conservation of global energy and local charge are important for the long-term accuracy of the simulation and the taming of temporal \cite{chen11} and finite-grid  \cite{barnes21} instabilities prevalent in explicit PIC models when such resolutions are employed. 

Compared to previous implicit PIC studies, our model builds on developments presented in \cite{chen11} and the generalization to mapped meshes presented in \cite{chac13}, but includes new features that are particularly advantageous for the study of magnetic nozzles. Worth mentioning are the introduction of the 1D paraxial geometry, the magnetic mirror force term, a new particle injection algorithm, and the dynamic downstream electric field boundary condition that overcomes the issues in capturing infinite plasma expansions in finite domains found in earlier studies. The novel and efficient segment-based mover \cite{chen22}, which allows particles to travel several cells in a single suborbit, was also generalized for the first time to mapped meshes.

The complete model has been verified, first in a periodic magnetic mirror that showcases the action of the mirror force and the conservation properties of the algorithm, and later in a relevant magnetic nozzle study. The MN example was analyzed thoroughly, observing only  minor differences with previous literature results and demonstrating convergence and low sensitivity to {\em ad hoc} control parameters to preserve quasineutrality asymptotically.
The algorithm is shown to scale favorably with the implicit timestep (i.e., larger timesteps lead to faster simulation times), and linearly (i.e., optimally) with both the number of particles and the number of mesh points. The latter result is particularly important, since it improves markedly on earlier implicit PIC implementations which relied on particles stopping at cell faces for charge conservation, and as a result scaled quadratically with the number of mesh points in one dimension \cite{chen14}.  

\section*{Acknowledgments}

The authors would like to thank Jiewei Zhou and Eduardo Ahedo at Universidad Carlos III de Madrid and Guangye Chen and Oleksandr Koshkarov at Los Alamos National Laboratory for the insightful discussions and inputs.
PJ and MM were funded by the European Research Council (ERC) under the European Union’s Horizon 2020 research and innovation programme (grant agreement No 950466).
LC was funded by the Applied Mathematics Research program of the Department of Energy Office of Applied Scientific Computing Research, and the research was performed under the auspices of the National Nuclear Security Administration of the U.S. Department of Energy at Los Alamos National Laboratory, managed by Triad National Security, LLC under contract 89233218CNA000001.

\bibliography{ep2, others, references}
\end{document}